\documentclass[sigconf]{acmart}

\usepackage{booktabs} 
\usepackage{balance}
\usepackage{csquotes}
\usepackage[colorinlistoftodos,prependcaption,textsize=tiny]{todonotes}

\usepackage{color}
\usepackage{textcomp}
\usepackage{booktabs}
\usepackage{ccicons}

\newcommand{\eat}[1]{\ignorespaces}

\newcommand{\squishlist}{
  \begin{list}{$\bullet$}
   {
     \setlength{\itemsep}{0pt}
     \setlength{\parsep}{3pt}
     \setlength{\topsep}{3pt}
     \setlength{\partopsep}{0pt}
     \setlength{\leftmargin}{1.5em}
     \setlength{\labelwidth}{1em}
     \setlength{\labelsep}{0.5em} } }

\newcommand{\squishlisttwo}{
  \begin{list}{$\bullet$}
   {
     \setlength{\itemsep}{0pt}
     \setlength{\parsep}{0pt}
     \setlength{\topsep}{0pt}
     \setlength{\partopsep}{0pt}
     \setlength{\leftmargin}{2em}
     \setlength{\labelwidth}{1.5em}
     \setlength{\labelsep}{0.5em} } }

\newcommand{\squishdef}{
  \begin{list}{}
   {
     \setlength{\itemsep}{0pt}
     \setlength{\parsep}{3pt}
     \setlength{\topsep}{3pt}
     \setlength{\partopsep}{0pt}
     \setlength{\leftmargin}{1.5em}
     \setlength{\labelwidth}{1em}
     \setlength{\labelsep}{0.5em} } }

\newcommand{\squishend}{
   \end{list} }

\begin{document}

\copyrightyear{2018}
\acmYear{2018}
\setcopyright{iw3c2w3}
\acmConference[WWW 2018]{The 2018 Web Conference}{April 23--27, 2018}{Lyon,
France}
\acmBooktitle{WWW 2018: The 2018 Web Conference, April 23--27, 2018, Lyon, France}
\acmPrice{}
\acmDOI{10.1145/3178876.3186034}
\acmISBN{978-1-4503-5639-8/18/04}

\fancyhead{}
\settopmatter{printacmref=false} 

\title{Web-Based VR Experiments Powered by the Crowd}
\author{Xiao Ma\textsuperscript{1,2}, 
	Megan Cackett\textsuperscript{2}, 
    Leslie Park\textsuperscript{2}, 
    Eric Chien\textsuperscript{1,2}, 
    and Mor Naaman\textsuperscript{1,2}}
\affiliation{%
  \institution{
  	\textsuperscript{1}Jacobs Institute, Cornell Tech \hspace{1em}
    \textsuperscript{2}Cornell University
    }
	\{xm75, mac389, lp343, jc3256, mor.naaman\}@cornell.edu
}


\begin{abstract}
We build on the increasing availability of Virtual Reality (VR) devices and Web technologies to conduct behavioral experiments in VR using crowdsourcing techniques. 
A new recruiting and validation method allows us to create a panel of eligible experiment participants recruited from Amazon Mechanical Turk. 
Using this panel, we ran three different crowdsourced VR experiments, each reproducing one of three VR illusions: place illusion, embodiment illusion, and plausibility illusion. 
Our experience and worker feedback on these experiments show that conducting Web-based VR experiments using crowdsourcing is already feasible, though some challenges---including scale---remain.  
Such crowdsourced VR experiments on the Web have the potential to finally support replicable VR experiments with diverse populations at a low cost.
\end{abstract}


\maketitle

{\small
\subsection*{ACM Reference Format:}
Xiao Ma, Megan Cackett, Leslie Park, Eric Chien, and Mor Naaman. 2018. Web-Based VR Experiments Powered by the Crowd. In WWW 2018: The 2018 Web Conference, April 23---27, 2018, Lyon, France. ACM, New York, NY, USA, 11 pages. https://doi.org/10.1145/3178876.3186034}

\section{Introduction} \label{introduction}
  Virtual Reality (VR) technology holds many promises, among which is the potential to enable a major paradigm shift for conducting social and psychological experiments~\cite{blascovich2002immersive}.  
A ``Virtual Reality'' is a simulated or artificial environment that allows or compels the user to have a sense of being present in an environment other than the one they are actually in~\cite{Schroeder1996}.
Virtual Reality can potentially address long-standing methodological problems for human subject experiments, including limited realism in experimental stimuli and environments, lack of replication, and nonrepresentative samples~\cite{blascovich2002immersive}.

However, the vision of VR enabling a new experimental paradigm remains largely unachieved for many reasons.
Despite a proliferation of studies both \emph{using} VR (e.g., restorative effects of virtual nature settings~\cite{valtchanov2010restorative}) and \emph{about} VR (e.g., avatar's impact on negotiation~\cite{yee2007proteus}), VR experiments are still limited in number and scope because of high associated costs. 
The cost of conducting VR experiments is three-fold: development costs including software development and deployment, experimental costs including physical space and co-presence of the experimenter, and equipment costs~\cite{yee2006walk,yee2007proteus,rosenberg2013virtual,lee2016wobbly,schwind2017hands}.
First, there is currently a steep learning curve for \emph{developing} VR applications, requiring special technical knowledge with game engines (e.g., Unity, Unreal Engine) and 3D modeling software, often device-specific. 
Second, experiments have traditionally required a dedicated VR lab for setting up and using VR equipment. In our review, most VR experiments, with a few notable exceptions~(e.g., ~\cite{oh2016immersion}), required a dedicated physical lab space. Moreover, in most VR experiments the experimenter was physically co-present with the participant, constraining the number of eligible participants and preventing running more than one session at a time, even though in some cases it was unnecessary or undesired for the experimenter to be physically co-present with the participant.
Finally, at least until recently, VR experiments were limited by the high cost and limited general availability of VR devices~(e.g., the nVisor SX head-mounted display used in~\cite{yee2007proteus} is estimated to cost \$15,000 when the study was conducted).
Taken together, the high costs of conducting VR experiments particularly discourages replication studies. In our review of VR studies we did not find software, data, or logs that could be re-run or re-analyzed even in recent studies~\cite{lee2016wobbly,schwind2017hands}.

While VR experiments have been facing challenges, crowdsourcing methods have exceedingly been used to conduct online experiments, especially on Amazon Mechanical Turk (AMT). 
Although the platform was originally built for human computation tasks~\cite{little2010turkit}, researchers have shown that AMT is a valid environment for conducting behavioral research~\cite{mason2012conducting} and other types of experiments~\cite{mao2016experimental,paolacci2010running,suri2011cooperation}. According to Mason and Suri, AMT allows for ``easy access to large, stable, and diverse subject pool, the low cost of doing experiments, and faster iteration between developing theory and executing experiments''~\cite{mason2012conducting}. Indeed, AMT has become a popular research tool for conducting behavioral experiments in areas like contagion~\cite{suri2011cooperation}, cooperation~\cite{mao2017resilient}, and teamwork~\cite{mao2016experimental}. At the same time, crowdsourced tasks have supported increased complexity: the state-of-art AMT research shows that crowdworkers are able to produce work where the output is comparable with domain-specific experts~\cite{chilton2013cascade}, as well as accomplish highly complex tasks such as app building~\cite{retelny2014expert} and fiction writing~\cite{kim2017mechanical}.

Can we use crowdsourcing techniques to run VR experiments on the Web? Such a proposition raises many questions and challenges. First, is there a large enough population of crowd workers with the devices needed to participate in VR experiments, and is this population not biased in key ways (e.g., socioeconomics)? Second, what are the technical VR platforms and environments available to developers, and what technologies are needed to deploy and run crowdsourced experiments? Third, can we develop VR experiments that can run remotely and independently---without an experimenter present, and without strict control over the physical environment and experiment execution?

This work addresses the questions and challenges raised above, making three key contributions. (1)~We develop a novel recruitment task where crowd workers on AMT can demonstrate access to a VR device. We survey that population of workers and show that there are hundreds of eligible workers of varied demographics with VR devices.~(2)~We implement and deploy three distinct VR experiments using AMT, and demonstrate their feasibility. These experiments were meant to replicate or mirror seminal studies in VR and behavioral research. In particular, the studies were chosen to reflect three types of illusions that VR is theorized to deliver: place illusion, plausibility illusion and embodiment illusion~\cite{gonzalez2017model}. Two out of the three studies replicated did not align with the original study's result; we highlight potential reasons for this outcome in the discussion. We make the complete code of the studies we deployed publicly available, allowing for easy adjustments and replication of our studies.~(3)~We use survey tools to understand the conditions and context of crowdsourcing workers performing VR tasks, and to reason about the constraints and considerations for VR AMT experiments done beyond lab settings. We show that, despite some technical issues and physical constraints, the deployments were mostly successful in providing VR-based ``illusions''. To the best of our knowledge, this is the first work to extend VR experiments to crowdsourcing platforms.

\section{Background} \label{background}
Our work builds on two major areas of research: (1)~running online experiments, especially using crowdsourcing techniques, and (2)~the growing set of behavioral research studies using Virtual Reality.

\subsection{Running Online Experiments}
There are many potential benefits for conducting experiments using crowdsourcing. According to Mason and Suri~\cite{mason2012conducting}, writing about one of the largest crowdsourcing services, Amazon Mechanical Turk (AMT), potential benefits include: (1)~subject pool diversity in terms of age, ethnicity, and socioeconomic status; (2)~low cost and built-in payment mechanisms; (3)~faster theory/experiment cycle; (4)~relative stability of the subject pool over time.
Paolacci et al. highlighted additional benefits of AMT crowdsourcing such as subject identifiability~\cite{paolacci2010running}, which we also leverage in this work. In AMT, crowdworkers can be required to earn ``qualifications'' by answering prescreening questions prior to participating in a study, allowing experimenters to have more control.
A major additional advantage is in the potential scale of distribution. Together with the often-lower cost, crowdsourcing allows for studies with larger numbers of participants. In turn, the larger scale provides an opportunity to apply more granular treatments~(e.g., a larger number of experimental conditions).

Several papers have supported the validity and effectiveness of crowdsourced experiments, in various fields, and in particular focusing on AMT. In~\cite{horton2011online, paolacci2010running}, the authors replicated experiments from game theory, heuristics, and biases literature using AMT, showing similar results to previous physical lab studies. Indeed, AMT has been a popular tool for behavioral experiments~(e.g.,~\cite{eriksson2010emotional,mao2016experimental,suri2011cooperation}), and it is important to continue to expand the capabilities of crowdsourced experiments.

A significant body of work addressed early challenges in crowdsourced experiments~\cite{dai2013pomdp,sheng2008get}.
Mason and Suri provided a detailed guide on how to conduct behavioral research on AMT~\cite{mason2012conducting} and described solutions to common problems that researchers might face when executing research to ensure high-quality work.
More recent work, however, focused on enabling more sophisticated studies and experimentation on AMT. For example, Mao et al. created TurkServer, a platform that facilitates synchronous and longitudinal experiments on AMT~\cite{mao2012turkserver,mao2017resilient}.
In general, as we note above, AMT study designs have demonstrated increasing complexity in multiple contexts ~\cite{chilton2013cascade,kim2017mechanical,lasecki2011real,retelny2014expert}.

Online experiments are not limited, of course, to crowdsourced environments, and have long been an acceptable tool in behavioral research~\cite{kraut2004psychological, salganik2017bit}.
Some recent innovations, however, involve new mechanisms of recruitment, distribution and data collection, for example by volunteers (in return for feedback on performance)~\cite{reinecke2015lab}. 
While such mechanisms will be increasingly available for research in VR, our focus remains on online VR experiments via crowdsourcing on AMT. With this in mind, next we review previous non-crowdsourced experiment research in VR.

\subsection{Virtual Reality and Behavioral Experiments}
A number of work hinted at conducting VR studies using crowdsourcing. Gehlbach et al. replicated in 2015 an earlier study on perspective taking conducted in VR~\cite{yee2006walk} using AMT---though the researchers replaced VR with Web-based 2D simulation~\cite{gehlbach2015many}. In 2016, Oh et al. proposed the concept of Immersion at Scale, testing out collecting data on mobile VR devices outside of the lab by setting up physical tents at different locations (e.g., at local events, museums)~\cite{oh2016immersion}. Most recently, researchers conducted the first ethnographic study in VR with remote participants~\cite{shriram2017all}. Our research takes the next step by executing real crowdsourced VR experiments. 

Outside of crowdsourced settings, there has been a proliferation of VR-based experiments in laboratory settings since Blascovich et al.'s call for using VR as a research tool in 2002~\cite{blascovich2002immersive}. We provide a short overview of some areas of experimental work, organized according to a taxonomy of VR ``models of illusion''~\cite{gonzalez2017model}. We also used this taxonomy to select the experiments we execute in this work.

In their taxonomy, Gonzales-Franco and Lanier argue that VR is capable in delivering primarily three types of illusions: place illusion, embodiment illusion, and plausibility illusion~\cite{gonzalez2017model}. Place illusion refers to a user's feeling of being transported into the rendered environment. Embodiment illusion refers to a user's feeling of experiencing the virtual world through an avatar. Together, place and embodiment illusions enhance the plausibility illusion, which refers to the feeling that events happening in the virtual world are real. Each type of illusion leads to different results. In general, researchers have been leveraging all three types of illusions in their studies to deliver different experimental manipulations.

\subsubsection{Place Illusion -- VR Environments} \label{placeIllusion}

It is well known that environments can impact various behaviors. There are two ways researchers can study the effect of environments: observation (e.g., reflective survey~\cite{dul2011knowledge}, camera capture~\cite{rawassizadeh2011persuasive}) and experimentation. In experiments, environment manipulations were achieved through different ways: physically bringing participants to a desired environment~\cite{hartig1991restorative}, physically bringing participants to a manipulated environment~\cite{vohs2013physical}, or showing participants photographs~\cite{sitton1984messy} or videos~\cite{ulrich1991stress} of desired or manipulated environments.

Place illusion naturally fits into the progression of experiment manipulation of environments---increasing in realism compared to photographs or videos. In fact, some studies have already used VR to deliver place illusion by investigating the effect of environments on behavior in VR. For example, Maani et al. showed immersion in cooling virtual environments during surgical procedures can reduce perceived pain levels~\cite{maani2011virtual}.
Emmelkamp et al. showed that exposure to virtual environments is as effective as exposure to real environments on reducing anxiety and avoidance of patients suffering from acrophobia~\cite{emmelkamp2002virtual}.
And finally, twenty years after Hartig et al. brought students to a real hike to study the restorative effects of natural environments~\cite{hartig1991restorative}, researchers put participants in a virtual forest to study the same effects in VR~\cite{valtchanov2010restorative}.

\subsubsection{Embodiment Illusion -- Avatars}
In many VR technologies, users can experience the virtual world through an avatar---a virtual representation of self. 
Compared to 2D computer or mobile avatars, VR technologies offer a higher degree of embodiment, hence inducing the embodiment illusion.

Many studies using VR technologies have demonstrated the influences of embodied experiences on behavior. 
Research has shown that it is possible in VR to generate perceptual illusions of ownership over a virtual body seen from first person perspective, and learn to control the virtual body~\cite{won2015homuncular} even when the body appears different from the user's real body~\cite{bergstrom2016firstperson}. In addition, different designs of avatars were also shown to impact the perceived levels of presence~\cite{schwind2017hands} and other behaviors.

In particular, one well known phenomenon in VR regarding the embodiment illusion is the \emph{Proteus effect}~\cite{yee2007proteus,yee2009proteus}. The Proteus effect refers to the phenomenon that characteristics of a user's virtual avatar influence the user's behavior.
For instance, Yee and Bailenson showed that participants assigned with more attractive avatars behaved more intimately with confederates in self-disclosure and interpersonal distance tasks, and participants assigned with taller avatars behaved more confidently in a negotiation task~\cite{yee2007proteus}.

Although widely cited, subsequent studies have found mixed evidence in supporting the Proteus effect. Additional support of the Proteus effect was exhibited in a study in which, the embodiment of sexualized avatars elicited higher reports of self-objectification~\cite{fox2013embodiment}. However, in another study on dyadic communication, the data collected was not consistent with the hypothesis derived from the Proteus effect~\cite{van2013proteus}. Researchers suggest further studies are needed to identify the boundary conditions of Proteus effect.

\subsubsection{Plausibility Illusion -- Transformed Realities}
Finally, plausibility illusion builds upon place and embodiment illusions, generating a feeling that events happening in the virtual world are real. Previous research has leveraged plausibility illusion to induce empathy: a study simulating red-green color blindness~\cite{ahn2013effect} led to higher willingness to assist color blind individuals; embodying superhero abilities~\cite{rosenberg2013virtual} led to greater instances of prosocial behavior; and experiencing aging of one's avatar led to a decrease in the stereotyping of the elderly~\cite{yee2006walk}. Plausibility illusion is the ultimate advantage of conducting VR experiments, compared to delivering experimental manipulations in less immersive media such as photographs and videos.

Taken together, the environment, embodiment and plausibility illusions guide our selection of studies to first run in this work. We provide more details in later sections.

\subsubsection{Scale and Samples in VR experiments}
An informal literature review of VR experiments suggests that the studies, while sometimes reaching a significant size, were indeed constrained by cost, lab availability, and the number of diverse participants. Except for the few notable exceptions above, all the studies we reviewed~(e.g.,~\cite{yee2006walk,yee2007proteus,rosenberg2013virtual,lee2016wobbly,schwind2017hands}) were performed in a physical lab, using the lab's VR equipment.
The scale of these experiments varied, including up to~158 participants in one study~\cite{oh2016let}. However, most experiments in our review included 25--90 participants. Moreover, the majority of the studies sampled participants from the local university population, i.e., 18--24 years old students (e.g.,~\cite{yee2007proteus,valtchanov2010restorative,won2015homuncular,rosenberg2013virtual}). Next, we provide an overview of the state of VR equipment and development environments, which motivates and enables crowdsourced experiments in VR that can expand scale and speed, reduce cost, and increase population diversity in VR experimentation.
\section{Virtual Reality Technologies}
The availability of hardware and software for various VR platforms has allowed the general reach of commercial VR devices to grow rapidly.
VR has three core components: a head-mounted display (HMD), a computing platform, and the sensors/controllers. 
There are two ways to categorize VR devices, depending on: if the device is room-scale or stationary, or if the computing platform is PC, mobile, or standalone. Room-scale VR allows users to freely walk around in the play area, while their real-world physical movements are tracked by dedicated sensors and reflected in the VR environment. Stationary VR, on the other hand, allows users to navigate through a virtual space using a controller, while users themselves physically remain at a static position. Stationary VR is usually cheaper than room-scale VR. PC/mobile VR refers to devices that require an attached computer or a mobile phone to perform computing on, while standalone VR does not require any additional hardware and all computing is performed on device.

Consumer VR devices, primarily in the PC/mobile VR category, began appearing on the market in 2014. Major competitors include: Google Cardboard (released Jun 2014, \$15, mobile, all prices are as advertised by maker on February 2018), Samsung Gear VR (Nov 2015, \$39.99, mobile), Oculus Rift (Mar 2016, \$399, PC), HTC Vive (Apr 2016, \$599, PC), Sony PlayStation VR (Oct 2016, \$399, PC), and Google Daydream View (Nov 2016, \$79, mobile). In terms of market share, Google Cardboard leads, having shipped 10 million units since its launch in 2014~\cite{robertson2017google}. Samsung Gear VR follows with shipping 5 million since its launch, followed by Sony PlayStation VR (750k units, all statistics on the units shipped are aggregates as of 2016)~\cite{kamen2017samsung}. In this work we focus on the stationary and mobile VR devices, as noted below. 

On the developer side, there has traditionally been a steep learning curve for developing non-trivial VR applications. For example, most VR developers use the game engines Unreal Engine (C++-based) and Unity (C\#). However, with the emergence of WebVR\footnote{\url{https://webvr.info/}} standards since 2015, as well as new JavaScript-based frameworks like React VR (2017), developing for VR is becoming easier and faster. In this work we demonstrate the use of the React VR framework, as detailed below.

\section{Recruitment and Technical Setup}
\label{sec:recruitment}
To run our VR experiments we selected one of the most popular crowdsourcing platforms, Amazon Mechanical Turk (AMT). The first challenge we had to address was creating a panel of eligible workers with access to VR devices that could be used for the experiments. Our recruiting effort, then, aimed to create this ``VR-AMT Panel'', get a lower-bound estimate on the size of addressable VR device owners on AMT, and investigate the demographic characteristics of this set of workers. 
Recruiting workers with VR device is of course dependent on the type of devices required. 
We chose to collect data on the most popular devices that are available for casual VR users as reviewed above, namely Google Cardboard, Samsung Gear VR, Oculus Rift, HTC Vive, Sony PlayStation VR, and Google Daydream View, which we refer to as ``Survey Devices'' below.

To identify and characterize the potential of the VR-AMT Panel, we created a unique ``qualification'' task on AMT, asking workers to demonstrate that they have access to a VR device. To this end, our task asked  workers to upload a picture of their VR device. 
Furthermore, to prove their access to the device rather than to a picture of one, we asked for the picture to include a piece of paper with the hand-written last four digits of the worker's ID: easy to do and verify if the worker actually has access to a VR device, but hard to fake if not. In addition, we asked about how the device was acquired and general experiences of using the device to gather input on workers' level of expertise using VR systems. 
Finally, in the same task we asked workers for demographic information including age, gender, race, education level, occupation, location, urban/rural, and household income.

\begin{figure}[t]
\includegraphics[width=3.2in]{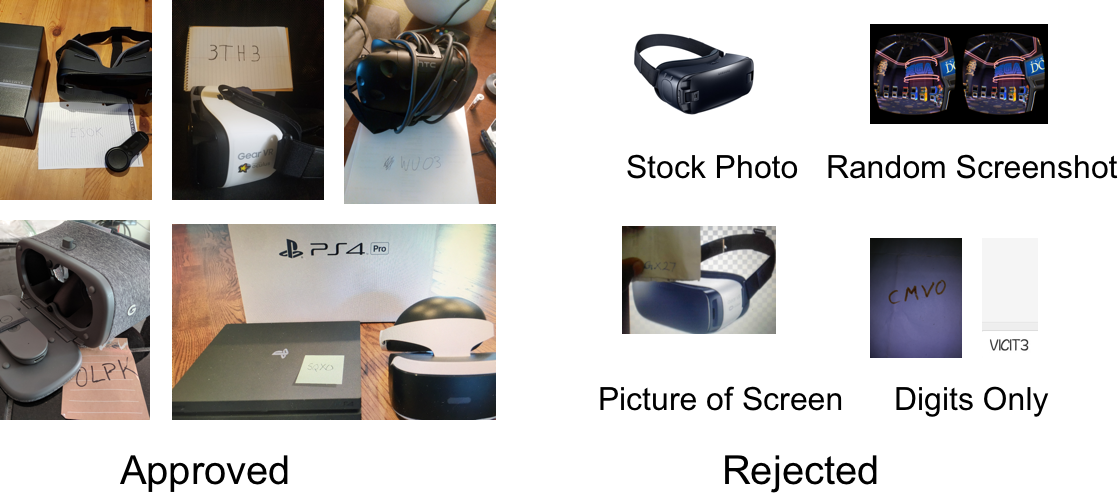}
\caption{Examples of approved and rejected validation image}
\label{screening}
\end{figure}

We launched the qualification task on AMT in multiple batches, compensating workers \$2 for an estimated work time of 10 minutes. The task description made clear that it is only open to people who have immediate access to VR devices listed in the Survey Devices set, or pre-approved by us through email communication. 
We accepted submissions for a period of 13 days in total, and received 439 submissions. We manually evaluated all the submissions to check whether workers followed the instructions, and accepted only submissions that included a valid picture with matching worker ID and a VR device from our Survey Devices set. Examples of valid and spam submissions are shown in~\autoref{screening}.

The outcome of this step was a VR-AMT Panel of 242 crowdsourced workers, with information on the type of VR devices they have access to, and their demographic information. We analyze this data next.

\subsection{Characteristics of VR-Eligible Workers}
\label{subsec:characteristics}
We analyzed the responses of the 242 valid qualification task submissions to reason about the demographics and characteristics of this addressable population of the VR-AMT Panel.  

\textbf{Demographics.}
The workers who made a valid submission of our qualification task do not differ significantly from previously reported demographics of AMT population in terms of age, gender, and household income~\cite{huff2015these,mason2012conducting,ipeirotis2010demographics}\footnote{The numbers are also similar to data collected by an AMT online tracker during the same period: \url{http://demographics.mturk-tracker.com/\#/gender/us}}.
Our sample was 61\% male, and ranged in age from 18 to 78 (sparkline: \raisebox{-.8ex}{\includegraphics[width=.3in]{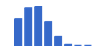}}), with a median age of 32. The sample was 70\% White, 14\% Asian,  and 6\% Black or African American, with 90\% United States residents. Out of the sample, 52\% were living in a suburban area, 30\% urban, and 18\% rural. Education ranged from less than a high school degree to a doctoral degree \raisebox{-.8ex}{\includegraphics[width=.23in]{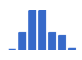}}, with the majority of workers having completed some college (30\%) or acquired a college-level bachelor's degree (30\%). Income levels varied from less than \$10k to over \$100k \raisebox{-.5ex}{\includegraphics[width=.3in]{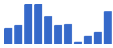}}, with a majority (56\%) earning between \$30k and \$80k~(the median US household income in 2016 was \$56k). The income characteristics show that access to VR devices was not limited to higher socioeconomic status. 

One difference between our panel and other AMT surveys~\cite{huff2015these,ipeirotis2010demographics} is the geographic location of workers. Our panel has a higher percentage of U.S.-based workers compared to others~\cite{ipeirotis2010demographics}, potentially because VR devices are more readily available in the U.S. market. The panel is also more suburban than reported in the general AMT population~\cite{huff2015these}. 
Still, the demographics of our validated panel of VR-eligible workers are much more diverse in terms of age, education level, occupation, and income than the WEIRD (Western, Educated, Industrialized, Rich, and Democratic) population that most previous VR studies draw from~\cite{yee2007proteus,valtchanov2010restorative,won2015homuncular,rosenberg2013virtual}.

\textbf{Worker VR history and usage.}
In the qualification task, we also asked the workers about their VR experience and use, with the goal of establishing the baseline level of worker VR skills.  
For brevity, we do not include the full results here, but note that most participants reported some use of VR apps, reporting ``basic'' to ``intermediate'' fluency for common interactions in VR, such as navigation, browsing the web, and typing. 

\subsection{Technical Platform and Environment Choice}

We decided to focus on Samsung Gear VR (Gear VR below) as the experimental platform for our crowdsourced VR experiments in this work. In our sample, 144 workers owned Gear VR, compared to 46 with Google Cardboard, 18 with HTC Vive, and 18 with Sony PlayStation VR. 
For simplicity of development and to avoid additional technical confounds, we chose to develop the experiments only for the Gear VR. 
Below, unless otherwise noted, when we refer to ``VR Panel'' of eligible workers we only include a subset of 122 validated Gear VR users (22 Gear VR users completed the qualification task after we launched our experiments). 

Reducing the sampling frame to Gear VR owners did not significantly change the demographics of workers. For example, our VR Panel population was 55\% male, compared to 61\% to the full VR-AMT Panel. 

To develop the experiment software we selected the React VR framework\footnote{Available from \url{https://facebook.github.io/react-vr/}}, a tool to ``compose a rich VR world and UI from declarative components'' using Javascript. 
React VR is somewhat more accessible than other development platforms such as Unity, and could more easily be adopted by researchers with lighter development experience. Moreover, React VR code is easier to understand and adapt, even for non-professional developers, easing the replication or adaptation task. We have open-sourced the code used in our experiment, along with our experimental data and logs, to allow for transparency and easy replication.\footnote{Available from \url{http://github.com/sTechLab/VRCrowdExperiments}}

\section{Study Overview}

We conducted three experimental studies in Virtual Reality on Amazon Mechanical Turk. The studies all followed a similar flow; we recruited participants from the VR Panel, provided study instructions, ran the experiment, and collected feedback in an exit survey.

\begin{figure}
\includegraphics[width=3in]{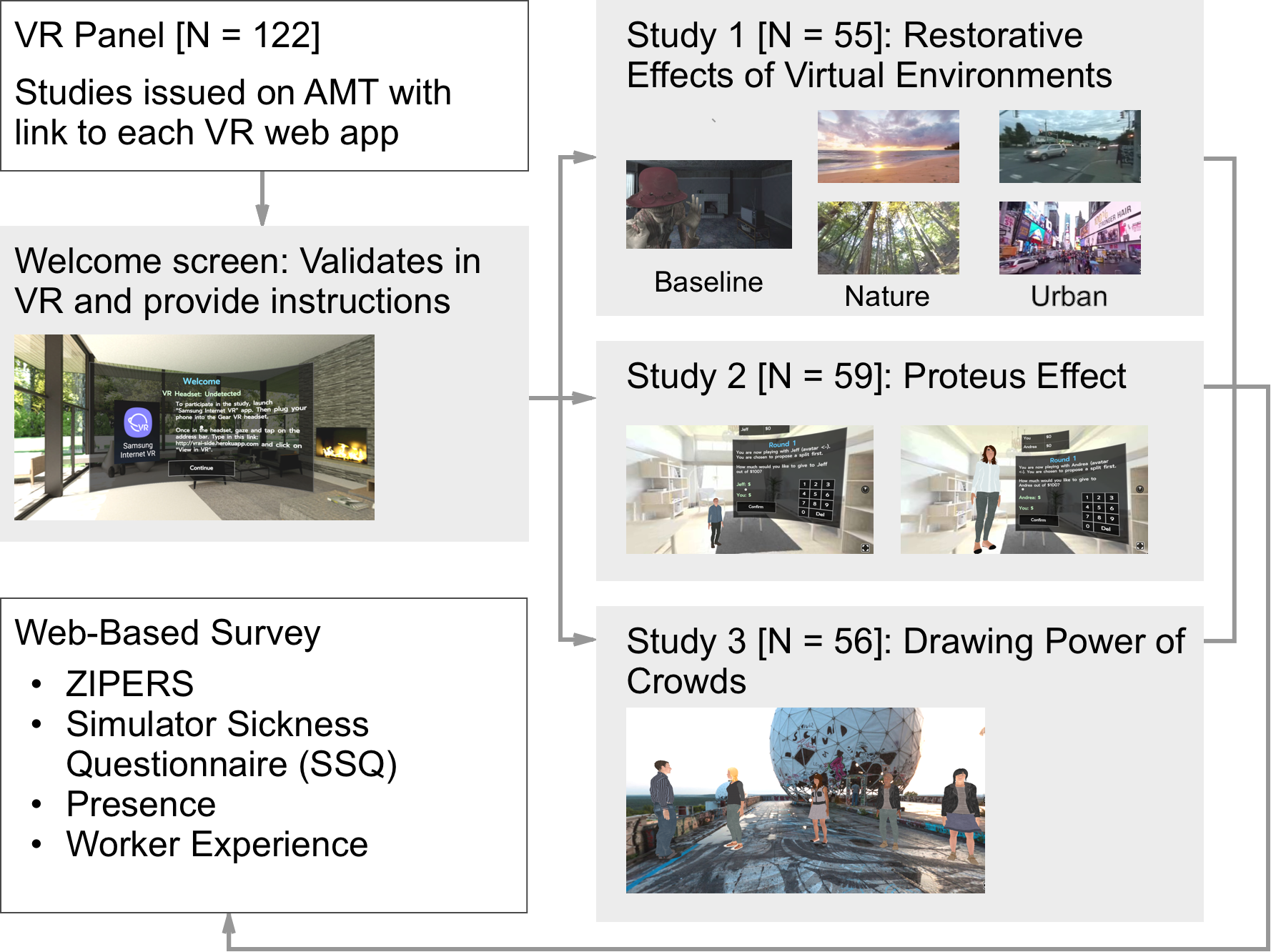}
\caption{Flow of experimental setup (sections with grey background were completed in VR)}
\label{flowchart}
\vspace{-10px}
\end{figure}

The general flow of the studies is shown in~\autoref{flowchart}. For each study, we issued a task on AMT with detailed instructions on what the workers should expect to see in the study. The task included a VR study link and the worker was instructed to navigate to the link while wearing the VR headset using a VR browser that supports WebVR standards (Samsung Internet or Oculus built-in browser). Our React VR Web app automatically checked if the participant was wearing the headset, and only enabled the ``continue'' button when a headset was detected. The Web app randomly assigned the participant to an experimental condition. The participant was presented with a brief introduction before proceeding to the main part of the VR experiment. After the VR experiment, the participant received a VR verification code that unlocked the final Web-based exit survey, available from the original task description on AMT. 
The exit survey included the standard simulator sickness questionnaire~\cite{kennedy1993simulator}, presence questionnaire~\cite{witmer1998measuring}, questions about the worker's experience, and additional variables we collected for each specific study.

\section{Study 1: Restorative Effects of Virtual Environments}

Our first study is centered around the ability of VR to deliver place illusion~\cite{gonzalez2017model}, i.e., the illusion of, ``being in a place in spite of the sure knowledge that you are not there''~\cite{slater2009place}.
Specifically, we investigate the restorative effects of virtual nature and urban environments, an established research problem with a history of research in environmental psychology showing the restorative effects of nature~\cite{ulrich1981natural,ulrich1991stress, valtchanov2010restorative}, urban~\cite{karmanov2008assessing}, and built~\cite{packer2010museums} environments. As noted in Section~\ref{placeIllusion}, research methods for studying the effect of environments have advanced from taking participants to the real physical environment~\cite{hartig1991restorative}, to using photographs~\cite{sitton1984messy} and videos~\cite{ulrich1991stress} as stimuli. The ability of VR to deliver place illusion is a natural progression, explored by various studies in lab settings~\cite{emmelkamp2002virtual,valtchanov2010restorative,maani2011virtual}. Here, for the first time, we evaluate whether such illusion can be delivered via crowdsourcing, while the experimenters have less control over the participant experience.

Our experiment is inspired by the VR study design of Valtchanov et al.~\cite{valtchanov2010restorative}, but differs in several important aspects. The original study first uses a math test as stressor to raise negative and decrease positive psychological states, as measured using the Zuckerman Inventory of Personal Reactions (ZIPERS) scale~\cite{zuckerman1977development}. Then, in a nature experimental condition, the study exposes participants to an active exploration of a virtual forest, while a control condition features an abstract slideshow in VR. Finally, the participants were measured again using the ZIPERS scale. The study used 22 participants and showed that the nature condition resulted in higher restorative effects than the abstract slideshow. 

We made some key adjustments in our experiment. First, our treatments included nature versus urban~(rather than abstract setting), following non-VR work~
on the topic~\cite{hartig1991restorative,ulrich1991stress,karmanov2008assessing}. Previous research predicts that the virtual nature environment will result in an increase in the participant's positive affect and a decrease in the negative affect (sadness, fear, anger) compared to the urban environment, as measured by the ZIPERS scale~\cite{zuckerman1977development}.
Second, since the math test of Valtchanov et al.~\cite{valtchanov2010restorative} was reported \textit{not} to be effective in raising negative affect level, we instead used a stress-inducing video in VR---a 360 trailer of a thriller movie. 
Instead of using the between-subject repeated-measures design of the original study~(measuring the ZIPERS scale twice for each participant), we used a similar between-subject design but measured ZIPERS only once at the end, to not interrupt the VR experience. 
In a baseline condition, participants proceeded directly to report their psychological states using the ZIPERS scale without viewing any ``restorative'' video. This condition acted as a manipulation check, showing whether the thriller video was indeed causing an increase in negative affect.

In our experiment, the nature and urban videos were selected to resemble closely with the footage described in~\cite{ulrich1991stress}. The nature video included nature vegetation and water, with sounds of birds, breeze, and waves. The urban video included scenes with 
light to heavy urban traffic with pedestrians and sounds of cars, voices, footsteps, and other people noises. 
~\autoref{flowchart} includes screenshots of the videos used for the study.

\subsection{Execution and Results}
We launched the VR experiment on Amazon Mechanical Turk, restricting the task such that only workers from our VR Panel could take it. The study task was available on AMT for seven days, and was removed from AMT when submissions stopped. In total, 66 workers performed this study task on AMT (whom we refer to as ``participants'' below). We paid each participant \$5, for a total cost of \$396 for the study after AMT fees. 
We filtered out participants who did not start the ZIPERS survey within twenty minutes of the VR portion, leaving us with 55 valid submissions. The participants were randomly assigned to conditions, with a final breakdown of baseline (19 participants), nature (24), and urban (12) conditions. 

\begin{figure}[h]
\includegraphics[width=3.2in]{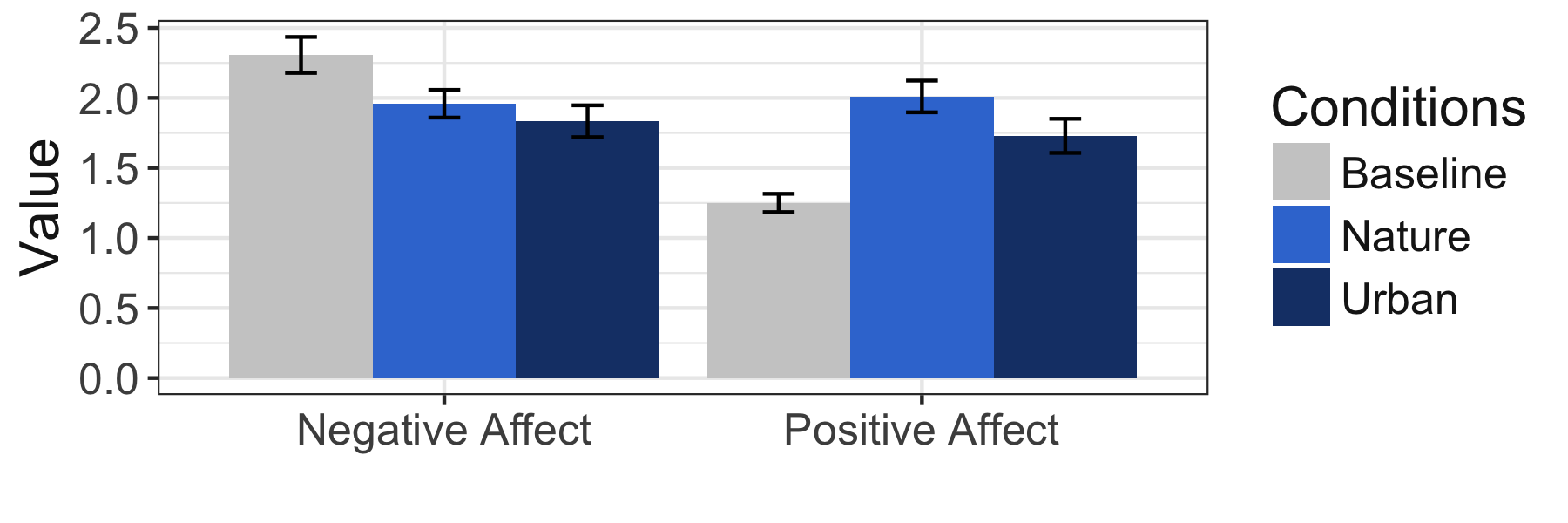}
\vspace{-15px}
\caption{Change in ZIPERS measures for different conditions. Error bars show +- standard error of the mean.}
\label{study1result}
\end{figure}

The results indicated potential restorative effects of both nature and urban settings, but showed no difference between the urban and nature conditions. 
We compared the positive affect and negative affect measured by ZIPERS among the two experimental conditions and the baseline. As shown in \autoref{study1result}, both the nature and urban conditions provided a reduction in negative affect and increase in positive affect compared to the baseline. An ANOVA analysis showed that this effect was significant for both positive ($p<.001$) and negative ($p<.05$) affect. 
A post-hoc Tukey analysis showed significant individual differences between the baseline and each of the urban and nature conditions in decreasing negative and increasing positive affects. All differences were significant at $p<.05$ or lower, except for a marginal difference between baseline and nature in negative affect ($p=.06$). There were no statistically significant differences between nature and urban conditions in all measures. We also note that ZIPERS included a focus measure, which was not expected to be impacted by the restorative treatments. Indeed, an ANOVA analysis did now show any difference of the focus measure across baseline and experimental conditions. 

\vspace{-18px}

\subsection{Discussion}
In this first study, we showed it is possible to run crowdsourced experiments in VR delivering place illusion. The study did not confirm the advantage of nature over urban environments for restorative effects. This outcome could be due to that the specific characteristics of the environment, beyond nature or urban make a difference, as hypothesized in other related studies~\cite{karmanov2008assessing}.
The different characteristics of environments  can be investigated further, for example, using our open-sourced code to conduct additional experiments.
At the same time, our stress-inducing treatment (VR thriller) was shown to be effective---in VR and AMT---and can be used in future studies. The fact that negative affect was reduced and positive affect increased after both treatments, compared to the baseline, shows that the manipulation was successful, providing further validation of the VR + AMT paradigm.


\section{Study 2: Proteus Effect}
In the second study, we examine the embodiment illusion via a study of the Proteus effect, a well-studied topic in VR~\cite{yee2007proteus,yee2009proteus,fox2013embodiment,schwind2017hands,van2013proteus}. The Proteus effect refers to the phenomenon that, ``individual's behavior conforms to their digital self-representation independent of how others perceive them''~\cite{yee2007proteus}. The original study that coined the term showed in one of the experiments that ``participants assigned taller avatars behaved more confidently in a negotiation task than participants assigned shorter avatars''~\cite{yee2007proteus}. We model our study after this experiment.  

The Proteus effect study manipulated the height of a user's avatar in VR and measured their confidence via the behavior in a negotiation task against another VR avatar operated synchronously by a confederate. The negotiation implemented was a version of the Ultimatum Game~\cite{nowak2000fairness}, in which a hypothetical pool of \$100 is split between the negotiating parties; one party chooses a split and the other would choose either to accept it (in which case, the money is shared accordingly) or to reject it (nobody would receive any money). 
Taller (in VR) and therefore more confident negotiators were hypothesized to suggest more skewed splits, and more readily reject unfair splits. 

We adapted our study in several ways. Running the study in AMT, we had to devise a way to run it without a confederate. Instead, we used a bot avatar that was programed to make specific bids and accept or reject offers according to consistent guidelines~(listed below). Additionally, we created a different manipulation of avatar height. Instead of showing both the user's avatar and the other~(confederate) avatar of a different height, the user only saw the other~(bot) avatar, whose scale was manipulated to be smaller or larger as a proxy for height (see~\autoref{flowchart}).
Finally, while the original study had the participants always play against an avatar of an opposite gender, we had each participant play against two avatars in total~(one male and one female, both white, in randomized order).

In our task setup, each participant first received a Web-based tutorial about the Ultimatum Game, and were asked to pass two test rounds of the game to make sure they understand the rules. We then directed the participant to the VR interface. There they were shown a brief description of the study. 
We asked the participants a few questions to ``configure'' their \textit{own} avatar, before seeing the first opponent (bot) avatar. These configuration questions were meant to create the illusion that the other (bot) avatar is another participant who had also customized their presentation.
The participant then played one set of a four-round Ultimatum Game with the first opponent, proposing to split in the first and third round. Consistent with~\cite{yee2007proteus}, the bot avatar was programmed to always accept a split if the amount proposed to give the avatar is equal or more than \$20. The avatar was also programmed to offer 50-50 and 25-75 split in favor of the avatar in the second and fourth round.  
At the completion of the first set of rounds, the same procedure was repeated for the second opponent for another four rounds. To support realistic play, we told participants they would get a bonus amount of \$1--5 for the task, depending on their rank in terms of total amount of money retained in the game.

\subsection{Execution and Results}
We launched the VR experiment on Amazon Mechanical Turk, again limiting participation to workers from our VR Panel. The study task was available on AMT for seven days, and was removed from AMT when submissions stopped. In total, 69 workers performed the task. We filtered out 10 submissions with incomplete data, retaining 59 submissions. We paid each participant \$5, for a total cost of \$582 after the bonuses and AMT fees. We note that 90\% of the participants chose an avatar of the same gender as the one they reported in the qualification task~(Section~\ref{sec:recruitment}).

There were two dependent variables in the study: (1)~the amount proposed by the participant to reserve for self as opposed to offer to the opponent in rounds 1, 3, 5, and 7; and (2)~the likelihood of the participant accepting an \textit{unfair} split from the avatars in round 4 and round 8. 
The average split offers by participants (around 60-40 in favor of self), as well as the likelihood to accept unfair splits (22\%) were comparable with rates reported in prior studies of the Ultimatum Game~\cite{yee2007proteus,yee2009proteus,nowak2000fairness}. 
A MANOVA analysis did not show a statistically significant difference with regard to splits \{1,3,5,7\} among different conditions even after filtering outliers following the original study~\cite{yee2007proteus}. 
Further, a Fisher's Exact Test analysis did not observe statistically significant differences in round~4 and~8 accepts between conditions.

\subsection{Discussion}
Based on the average offer splits and accept rates we conclude that most participants played the game in earnest. However, our experiment did not result in the same outcomes reported by the original work~\cite{yee2007proteus}.  
There are several potential explanations for this replication discrepancy. First, it is possible that our VR illusion was not accepted by workers. The consistent game play showed that the workers at least accepted the settings of the game, if not the illusion of embodiment. 
Second, the discrepancy could be due to the difference between the more diverse worker population and the original study's sample of college students. 
Third, our treatments~(scale) were different in the way that we expressed height in the VR interface.
Finally, it could be that the study hypothesis is simply not valid. Importantly, our study here could be easily modified and replicated to examine the hypothesis more broadly using our open-sourced code and experimental data.
\section{Study 3: Drawing Power of Crowds}
In the final study, we investigate the plausibility illusion, the ``illusion that what is apparently happening is really happening~(even though you know for sure that it is not)''~\cite{slater2009place}. 
Specifically, we demonstrate the plausibility illusion using a study not based in VR, but instead a classic field study in psychology:~Milgram et al.'s work on the drawing power of crowds~\cite{milgram1969note}. 
Replicating this study in VR would demonstrate how Web-based VR crowd experiments can unlock the ability to conduct fast and low-cost studies that mimic physical crowds.

The original study by Milgram et al. examines the drawing power of crowds of different sizes~\cite{milgram1969note}. In typical urban settings, if a group of people engage in an action simultaneously, they have the capacity to draw others into the crowd. 
In the original study, the authors employed a \emph{stimulus crowd}, which we replace with virtual avatars. 
The original study had stimulus crowds of different sizes staring up in the air in a public area, and then measured how many passer-bys looked up as a function of the size of this stimulus crowd.

In our replication, the passer-by was the VR-based participant, while different subsets of ten virtual avatars served as the stimulus crowd. Milgrams et al.'s orginal work required manual video-tapes to capture and annotate head orientations. In contrast, our dependent measure was automatically recorded and more granular. We access and record the participant's head orientation in 3D automatically and frequently through sensors in VR devices and React VR API. We made a slight modification to the original study: to study horizontal gaze (looking left and right, back and forth), rather than vertical gaze (looking up and down).

In the study, we constructed ten static avatars of various appearances, and artificially controlled the direction each avatar was facing.
The study had four conditions: zero, low, medium, and high, each with a randomly selected subset of 0, 2, 4, and 8 avatars facing the participant (``looking back''). 
We logged participant's head orientation inside the VR headset in degrees for pitch (looking left and right, back and forth) five times per second, and calculated the attention distribution for each condition.
We expect to see differences in the distributions of pitch position between conditions. Specifically, as the number of avatars looking ``back'' increases, we expect the draw of the crowd to nudge the participants to look outside of the default field of view (facing front), and align more with the crowd's gaze. 

\subsection{Execution and Results}
Again, we published the task on AMT and had the study open for seven days. We received 66 submissions in total. We paid each participant \$5, for a total cost of \$396 for the study after fees. After filtering for valid survey response, we retained 56 submissions, with 15, 15, 13, and 13 in each of the zero, medium, low, and high conditions respectively. 
In this study, the VR portion brought the participant into a large plaza where they could see the avatars (see Study 3 in ~\autoref{flowchart}). To make sure participants explored the scene, they were given an object-finding task with a time limit of three minutes. 
The participants were instructed that an animal-shaped object could appear at any given time and at a random location, and the object~(a 3D-animated fox) always appeared ten seconds before the end of the task, at a fixed position. 

\begin{figure}
\includegraphics[width=3.2in]{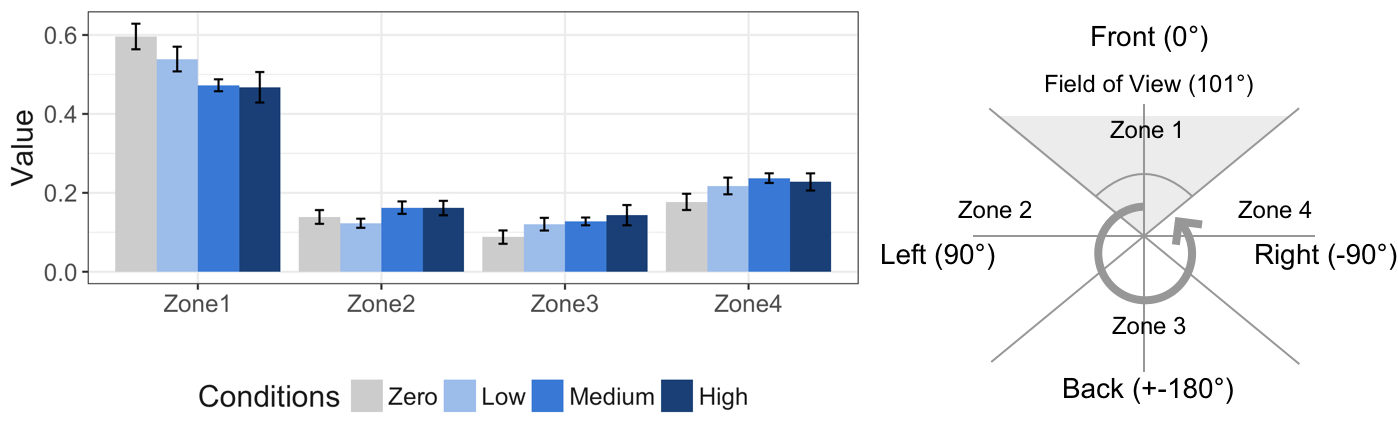}
\caption{Distribution of tracked head pitch positions by zone. Error bars show +- standard error of the mean.}
\label{study3result}
\vspace{-10px}
\end{figure}

We analyzed pitch positions of participants in each condition. In~\autoref{study3result}, we divide the pitch area to four zones, front, back, and sides numbered counter-clockwise 1--4. Gear VR has a field of view of 101 degrees, defining our front and back areas.  
~\autoref{study3result} shows a barchart of the portion of the experiment time participants spent in each area (zone~1 to~4), for each condition (zero to high, by color). 
~\autoref{study3result} shows that for the zero and low conditions, participants spent more time in zone~1 compared to medium and high conditions. Participants in medium and high conditions instead spent more time looking at zone 2--4. 
An ANOVA showed that the amount of time spent in zone~1 versus the rest was significantly different among different conditions ($p<.001$). A post-hoc Tukey test showed that in both medium and high conditions participants spent more time exploring \textit{outside} of zone~1 compared to both zero and low conditions ($p<.05$ in all cases). 

\subsection{Discussion}
The AMT-based VR experiment clearly demonstrated the virtual crowd's impact on the gazing behavior of a VR participant.
Importantly, this experiment demonstrated a key advantage of VR experiments---the ability deliver some accurate (though limited) measurements that are not available in other settings (online or offline). In particular, we used the head movement (approximating gaze) in this experiment to make the outcome more robust than a similar offline study could have achieved at a lower cost.

\section{Worker Experience}
\label{sec:experience}
A key goal of our study is to understand the experiences of workers participating in online VR experiments. To this end, for each study, we asked workers to complete an exit survey. In the survey, we asked about workers' experiences in VR environments including: level of simulator sickness and presence; the physical environments they were in; and any technical difficulties they may have encountered. We also included open-ended questions soliciting feedback from workers on general challenges they encountered.

\subsection{Workers' Experience in the VR Environment}
The simulator sickness (measured by SSQ~\cite{kennedy1993simulator}, $M$=$12.69$, $SD$=$14.71$; $M$=$10.12$, $SD$=$14.72$; $M$=$7.33$, $SD$=$11.4$ for three studies each) and the level of presence (measured by a shortened version of~\cite{witmer1998measuring}, $M$=$5.06$, $SD$=$.75$; $M$=$3.77$, $SD$=$1.05$; $M$=$3.93$, $SD$=$.87$) reported by the workers were not unsimilar to measures reported in previous VR studies~\cite{kennedy1993simulator,hock2017carvr,cciflikli2010increasing}. Study~1 stood out as generating both higher presence and higher simulator sickness rating, potentially due to the more engaging and realistic environment rendered by video.

In terms of general experience, one concern, raised by a few workers was the switching between VR and Web interfaces (\autoref{flowchart}). Common problems included mistyping or misremembering the VR link or the verification code. Future crowdsourced VR experiments could potentially improve the flow between VR and the Web, though it is not immediately clear how to do that on AMT. 

VR illusions often require full immersion of multiple senses. In Study~1, we requested the workers to wear earphones and have sound on. In the exit survey, we asked whether workers complied with the instructions. The survey clearly stated that workers would not be penalized for answering truthfully. The workers reported having earphones and sound on in 80\% of the cases, with an additional 15\% reported having sound on with speakers on the phone. 
These results suggest the workers are likely to follow instructions including sound that would lead to full VR immersion. 

\subsection{Workers' Physical Environments}
As we transit VR experiments out of the lab and into the worker's chosen physical space, we asked workers to provide details about the environment in which they were performing the VR tasks. Understanding different worker environments and potential interruptions is key to understanding whether such environments could support maintained VR illusions.

Across three studies, a high proportion---98\% of workers---reported performing the VR task at home. 
In their homes, workers reported a mixture of environments, most commonly: living room (24\%), bedroom (18\%), and home office (18\%). The majority of workers~(84\%) reported being alone; the rest reported the presence of one other person (14\%) or two other people (2\%) in the room as they were performing the task. Overall, then, we conclude most workers were in a physical environment that allowed them to complete the tasks uninterrupted by others.

We also asked workers about the dimensions of their physical environment to understand the potential of performing VR experiments that require body movement. The majority of workers reported having ample space: around 81\% of the workers reported having ``enough space to walk around'', and 10\% reported having the space to ``run around''. Only 6\% workers reported being limited in standing up or moving around.
At the same time, most workers~(89\%) reported being seated during the study, and the rest were standing. More than half (58\%) of the workers reported using a swivel chair that can rotate 360 degrees, with the remaining using a sofa/couch (13\%), stationary chair (11\%), bed (4\%), or other (14\%). In summary, the worker environments, as reported, could allow for experiments that require limited movement, though clearly we do not expect to be able to completely mimic lab-like settings with advanced body tracking capabilities and safety assurance.

\subsection{VR Devices and Technical Setup}
Beyond the physical setup, a challenging aspect of crowdsourced VR experiments in the wild may be the differences in devices used and technical setup; for example bandwidth limitations. To better understand these concerns, we asked the workers about any technical difficulties they may have encountered.  

Overall, workers used the Samsung Galaxy S7 (68\%), S6 (11\%), S8 (8\%), and other Samsung devices (e.g., Note 4/5) with the Gear VR headset. Most workers (95\%) did not report many technical issues.
Three workers reported issues of overheating and battery exhaustion. 
There were also seven reports of lagging rendering with fast head movements. Although, overall workers reported low levels of lagging---with a mean of $1.86$ ($SD$=$1.07$) on the scale of~1 (not at all) to~5 (a great deal). 
Related to these concerns, eight workers reported noticing graphic glitches, such as low quality image/video assets, or the clipping and flickering of avatars.

\section{Discussion}
\label{sec:discussion}
Reflecting on our studies, we can offer insight regarding the potential and drawbacks of crowdsourced VR experiments on the Web. 
First, our findings suggest that \textit{there is an addressable population of workers with VR equipment that is fundamentally more diverse than the ``regular'' VR studies' participants}, though (currently) limited in scale. The demographics of our study participants, as well as the demographics of the full panel of workers who have access to VR devices, had age, gender and socioeconomic diversity on par with the general AMT population.
These populations are likely to be significantly more diverse than the Western, Educated, Industrialized, Rich, and Democratic population used in most VR studies to date.

At the same time, the sampling frame for VR studies in AMT is still limited. We identified 242 workers with access to VR devices, and used a panel of 122 workers with Samsung Gear VR as the sampling frame for our work. Indeed, these numbers were large enough to reach a final sample of about 60 participants for each study~(with on average 85\% overlap rate of participants across studies). 
Moreover, extending the sampling frame to the 242 workers with verified devices would allow current experiments to reach---assuming 50\% response rate---a size comparable to the largest VR studies performed to date. 

While these numbers are promising, such scale is not yet sufficient to run unconstrained VR experiments using AMT. For example, such numbers do not yet allow to have more than a few conditions in each experiment (our experiments above ranged from two to four conditions) and limit the power of analysis based on other variables (such as gender in Study~2).
Moreover, with such scale, replication and re-testing can present an issue. We cannot easily sample ``fresh'' participants from the worker pool for a follow-up experiment. On the other hand, Paolacci et al.~\cite{paolacci2010running} points out that worker identifiability presents an opportunity for longitudinal studies, collecting data from the same group of people over time. 
Regardless, the growth of the consumer VR market may suggest that a larger VR-eligible pool of workers may be available in the future on AMT and other online platforms.

Additionally, our findings indicate that \textit{VR illusions can successfully be transmitted over the Web via crowdsourcing platforms using market-available devices and with low experimental cost}, even though the market-available devices limit the type of manipulations and the different measurements that can be done in crowdsourced VR.

In our AMT-based VR experiments, participants used the widely-available Samsung Gear VR headset. Such equipment does not allow hand movements tracking nor physiological measurements, like heart rate, muscle tension, etc., that were used in other VR studies~\cite{ulrich1991stress}. In addition, as detailed in Section~\ref{sec:experience}, the Samsung mobile phones used to power the headset have performance limitations---making the entire setup more fit for VR experiences that are not high-bandwidth, and do not require high-resolution rapid rendering. As VR devices continue to improve in quality and drop in price, it is reasonable to expect that more crowdworkers will have access to better devices that will allow more sophisticated experiments.

Despite the limitations, we have several indications that we were able to deliver effective VR illusions in the crowdsourced setting. First, the worker participants reported levels of presence on par with other studies~\cite{witmer1998measuring,cciflikli2010increasing} and reported few issues with the technical setup (Section~\ref{sec:experience}). Moreover, while not every one of our studies replicated the original results (more on that below), each of our studies included at least one successful manipulation: inducing stress in Study~1, the split offers consistent with prior work in Study~2, and the stimulus crowd (of virtual agents) in Study~3. 

Compared to lab studies, the crowdsourcing setup significantly reduces the experiments' direct costs. The total cost of running the combined 201 participants in our experiments was \$1,374, plus \$292.8 for the recruitment of the VR Panel. Compared to lab studies, our setup does not require the variable experimenter costs; reduces the time required from participants and therefore their associated costs; does not have lab space costs; and does not require device costs. Indeed, we estimate the total lab-based cost for running the exact studies described in the paper to be around \$3,600.\footnote{Based on 200 participants, estimated 30 minutes experiment duration including overhead time, \$10/hr wages for participants and experimenter, \$600 equipment costs and \$1000 lab space cost for 100 hours of use} 

Of course, the physical setup of crowdsourced VR was much more limited compared to a controlled and dedicated physical lab. This limitation may impact the quality of the VR illusion in crowdsourced studies. Additional manipulation or tracking mechanisms would need to be implemented to verify that workers complete studies that required specific physical action, e.g. standing~\cite{yee2007proteus}.

Finally, we do note that two of the three studies we executed in this work did not confirm the original study hypotheses. There could be many reasons for such outcomes, including deficiencies in the VR illusion. However, for the reasons stated above, we do not believe that was the case. More likely explanations are that (1)~the hypotheses were not justified, or (2)~the manipulations we had chosen were not effective. To fix the latter, anyone could download and fork our open-sourced code, and launch their own AMT-based VR replication study at a minimal time and monetary cost.\footnote{Available from \url{http://github.com/sTechLab/VRCrowdExperiments}}

\section{Conclusion and Future Work}
In this work, we have demonstrated the feasibility of running crowdsourced Virtual Reality experiments on the Web. Such experiments have the potential to finally support replicable VR experiments, with diverse populations, at a low cost and high speed. The number of available workers with VR devices continues to grow, and will allow the scale of such experiments to expand as well. Of course, expanding the experiments beyond AMT and even beyond crowdsourcing platforms using other online experiment recruiting methods could also be feasible and scalable~\cite{reinecke2015lab}.

Future work can build on ours to advance the type of experiments that are done in VR in to more complicated experimental setups. For example, future work could mirror previous (non-VR) crowdsourcing work to enable real-time~\cite{lasecki2011real, bernstein2011crowds}, synchronous~\cite{gordon2015legiontools}, or longitudinal~\cite{mao2012turkserver} experiments in VR.

\section{Acknowledgments}
We thank the crowdworkers who participated in our studies, and Oculus for providing the equipment used for developing the studies.  
We thank Dan Goldstein, Jake Hofman and Sid Suri for early feedback and direction. 
This work is partially supported by Oath through the Connected Experiences Lab at Cornell Tech.

\balance
\bibliographystyle{ACM-Reference-Format}
\bibliography{bib/myref,bib/vr,bib/lab} 


\begin{thebibliography}{62}


\ifx \showCODEN    \undefined \def \showCODEN     #1{\unskip}     \fi
\ifx \showDOI      \undefined \def \showDOI       #1{#1}\fi
\ifx \showISBNx    \undefined \def \showISBNx     #1{\unskip}     \fi
\ifx \showISBNxiii \undefined \def \showISBNxiii  #1{\unskip}     \fi
\ifx \showISSN     \undefined \def \showISSN      #1{\unskip}     \fi
\ifx \showLCCN     \undefined \def \showLCCN      #1{\unskip}     \fi
\ifx \shownote     \undefined \def \shownote      #1{#1}          \fi
\ifx \showarticletitle \undefined \def \showarticletitle #1{#1}   \fi
\ifx \showURL      \undefined \def \showURL       {\relax}        \fi
\providecommand\bibfield[2]{#2}
\providecommand\bibinfo[2]{#2}
\providecommand\natexlab[1]{#1}
\providecommand\showeprint[2][]{arXiv:#2}

\bibitem[\protect\citeauthoryear{Ahn, Minh Tran~Le, and Bailenson}{Ahn
  et~al\mbox{.}}{2013}]%
        {ahn2013effect}
\bibfield{author}{\bibinfo{person}{Sun Joo-Grace Ahn}, \bibinfo{person}{Amanda
  Minh Tran~Le}, {and} \bibinfo{person}{Jeremy Bailenson}.}
  \bibinfo{year}{2013}\natexlab{}.
\newblock \showarticletitle{The Effect of Embodied Experiences on Self-Other
  Merging, Attitude, and Helping Behavior}.
\newblock   \bibinfo{volume}{16} (\bibinfo{date}{01} \bibinfo{year}{2013}),
  \bibinfo{pages}{7--38}.
\newblock


\bibitem[\protect\citeauthoryear{Bergstr\"om, Kilteni, and Slater}{Bergstr\"om
  et~al\mbox{.}}{2016}]%
        {bergstrom2016firstperson}
\bibfield{author}{\bibinfo{person}{Ilias Bergstr\"om},
  \bibinfo{person}{Konstantina Kilteni}, {and} \bibinfo{person}{Mel Slater}.}
  \bibinfo{year}{2016}\natexlab{}.
\newblock \showarticletitle{First-Person Perspective Virtual Body Posture
  Influences Stress: A Virtual Reality Body Ownership Study}.
\newblock \bibinfo{journal}{\emph{PLOS ONE}} \bibinfo{volume}{11},
  \bibinfo{number}{2} (\bibinfo{date}{02} \bibinfo{year}{2016}),
  \bibinfo{pages}{1--21}.
\newblock
\urldef\tempurl%
\url{https://doi.org/10.1371/journal.pone.0148060}
\showDOI{\tempurl}


\bibitem[\protect\citeauthoryear{Bernstein, Brandt, Miller, and
  Karger}{Bernstein et~al\mbox{.}}{2011}]%
        {bernstein2011crowds}
\bibfield{author}{\bibinfo{person}{Michael~S. Bernstein}, \bibinfo{person}{Joel
  Brandt}, \bibinfo{person}{Robert~C. Miller}, {and} \bibinfo{person}{David~R.
  Karger}.} \bibinfo{year}{2011}\natexlab{}.
\newblock \showarticletitle{Crowds in Two Seconds: Enabling Realtime
  Crowd-powered Interfaces}. In \bibinfo{booktitle}{\emph{Proceedings of the
  24th Annual ACM Symposium on User Interface Software and Technology}}
  \emph{(\bibinfo{series}{UIST '11})}. \bibinfo{publisher}{ACM},
  \bibinfo{address}{New York, NY, USA}, \bibinfo{pages}{33--42}.
\newblock
\showISBNx{978-1-4503-0716-1}


\bibitem[\protect\citeauthoryear{Blascovich, Loomis, Beall, Swinth, Hoyt, and
  Bailenson}{Blascovich et~al\mbox{.}}{2002}]%
        {blascovich2002immersive}
\bibfield{author}{\bibinfo{person}{Jim Blascovich}, \bibinfo{person}{Jack
  Loomis}, \bibinfo{person}{Andrew~C Beall}, \bibinfo{person}{Kimberly~R
  Swinth}, \bibinfo{person}{Crystal~L Hoyt}, {and} \bibinfo{person}{Jeremy~N
  Bailenson}.} \bibinfo{year}{2002}\natexlab{}.
\newblock \showarticletitle{Immersive virtual environment technology as a
  methodological tool for social psychology}.
\newblock \bibinfo{journal}{\emph{Psychological Inquiry}} \bibinfo{volume}{13},
  \bibinfo{number}{2} (\bibinfo{year}{2002}), \bibinfo{pages}{103--124}.
\newblock


\bibitem[\protect\citeauthoryear{Chilton, Little, Edge, Weld, and
  Landay}{Chilton et~al\mbox{.}}{2013}]%
        {chilton2013cascade}
\bibfield{author}{\bibinfo{person}{Lydia~B Chilton}, \bibinfo{person}{Greg
  Little}, \bibinfo{person}{Darren Edge}, \bibinfo{person}{Daniel~S Weld},
  {and} \bibinfo{person}{James~A Landay}.} \bibinfo{year}{2013}\natexlab{}.
\newblock \showarticletitle{Cascade: Crowdsourcing taxonomy creation}. In
  \bibinfo{booktitle}{\emph{Proceedings of the SIGCHI Conference on Human
  Factors in Computing Systems}}. ACM, \bibinfo{pages}{1999--2008}.
\newblock


\bibitem[\protect\citeauthoryear{{\c{C}}iflikli, {\.I}{\c{s}}ler, and
  G{\"u}d{\"u}kbay}{{\c{C}}iflikli et~al\mbox{.}}{2010}]%
        {cciflikli2010increasing}
\bibfield{author}{\bibinfo{person}{Burak {\c{C}}iflikli},
  \bibinfo{person}{Veysi {\.I}{\c{s}}ler}, {and} \bibinfo{person}{U{\u{g}}ur
  G{\"u}d{\"u}kbay}.} \bibinfo{year}{2010}\natexlab{}.
\newblock \showarticletitle{Increasing the sense of presence in a simulation
  environment using image generators based on visual attention}.
\newblock \bibinfo{journal}{\emph{Presence: Teleoperators and Virtual
  Environments}} \bibinfo{volume}{19}, \bibinfo{number}{6}
  (\bibinfo{year}{2010}), \bibinfo{pages}{557--568}.
\newblock


\bibitem[\protect\citeauthoryear{Dai, Lin, Mausam, and Weld}{Dai
  et~al\mbox{.}}{2013}]%
        {dai2013pomdp}
\bibfield{author}{\bibinfo{person}{Peng Dai}, \bibinfo{person}{Christopher~H.
  Lin}, \bibinfo{person}{Mausam}, {and} \bibinfo{person}{Daniel~S. Weld}.}
  \bibinfo{year}{2013}\natexlab{}.
\newblock \showarticletitle{POMDP-based control of workflows for
  crowdsourcing}.
\newblock \bibinfo{journal}{\emph{Artificial Intelligence}}
  \bibinfo{volume}{202}, \bibinfo{number}{Supplement C} (\bibinfo{year}{2013}),
  \bibinfo{pages}{52 -- 85}.
\newblock
\showISSN{0004-3702}


\bibitem[\protect\citeauthoryear{Dul, Ceylan, and Jaspers}{Dul
  et~al\mbox{.}}{2011}]%
        {dul2011knowledge}
\bibfield{author}{\bibinfo{person}{Jan Dul}, \bibinfo{person}{Canan Ceylan},
  {and} \bibinfo{person}{Ferdinand Jaspers}.} \bibinfo{year}{2011}\natexlab{}.
\newblock \showarticletitle{Knowledge workers' creativity and the role of the
  physical work environment}.
\newblock \bibinfo{journal}{\emph{Human resource management}}
  \bibinfo{volume}{50}, \bibinfo{number}{6} (\bibinfo{year}{2011}),
  \bibinfo{pages}{715--734}.
\newblock


\bibitem[\protect\citeauthoryear{Emmelkamp, Krijn, Hulsbosch, De~Vries,
  Schuemie, and Van~der Mast}{Emmelkamp et~al\mbox{.}}{2002}]%
        {emmelkamp2002virtual}
\bibfield{author}{\bibinfo{person}{PMG Emmelkamp}, \bibinfo{person}{M Krijn},
  \bibinfo{person}{AM Hulsbosch}, \bibinfo{person}{S De~Vries},
  \bibinfo{person}{MJ Schuemie}, {and} \bibinfo{person}{CAPG Van~der Mast}.}
  \bibinfo{year}{2002}\natexlab{}.
\newblock \showarticletitle{Virtual reality treatment versus exposure in vivo:
  a comparative evaluation in acrophobia}.
\newblock \bibinfo{journal}{\emph{Behaviour research and therapy}}
  \bibinfo{volume}{40}, \bibinfo{number}{5} (\bibinfo{year}{2002}),
  \bibinfo{pages}{509--516}.
\newblock


\bibitem[\protect\citeauthoryear{Eriksson and Simpson}{Eriksson and
  Simpson}{2010}]%
        {eriksson2010emotional}
\bibfield{author}{\bibinfo{person}{Kimmo Eriksson} {and} \bibinfo{person}{Brent
  Simpson}.} \bibinfo{year}{2010}\natexlab{}.
\newblock \showarticletitle{Emotional reactions to losing explain gender
  differences in entering a risky lottery}.
\newblock \bibinfo{journal}{\emph{Judgment and Decision Making}}
  \bibinfo{volume}{5}, \bibinfo{number}{3} (\bibinfo{year}{2010}),
  \bibinfo{pages}{159}.
\newblock


\bibitem[\protect\citeauthoryear{Fox, Bailenson, and Tricase}{Fox
  et~al\mbox{.}}{2013}]%
        {fox2013embodiment}
\bibfield{author}{\bibinfo{person}{Jesse Fox}, \bibinfo{person}{Jeremy~N.
  Bailenson}, {and} \bibinfo{person}{Liz Tricase}.}
  \bibinfo{year}{2013}\natexlab{}.
\newblock \showarticletitle{The Embodiment of Sexualized Virtual Selves: The
  Proteus Effect and Experiences of Self-objectification via Avatars}.
\newblock \bibinfo{journal}{\emph{Computers in Human Behavior}}
  \bibinfo{volume}{29}, \bibinfo{number}{3} (\bibinfo{year}{2013}),
  \bibinfo{pages}{930--938}.
\newblock
\showISSN{0747-5632}


\bibitem[\protect\citeauthoryear{Gehlbach, Marietta, King, Karutz, Bailenson,
  and Dede}{Gehlbach et~al\mbox{.}}{2015}]%
        {gehlbach2015many}
\bibfield{author}{\bibinfo{person}{Hunter Gehlbach}, \bibinfo{person}{Geoff
  Marietta}, \bibinfo{person}{Aaron~M King}, \bibinfo{person}{Cody Karutz},
  \bibinfo{person}{Jeremy~N Bailenson}, {and} \bibinfo{person}{Chris Dede}.}
  \bibinfo{year}{2015}\natexlab{}.
\newblock \showarticletitle{Many ways to walk a mile in another's moccasins:
  Type of social perspective taking and its effect on negotiation outcomes}.
\newblock \bibinfo{journal}{\emph{Computers in Human Behavior}}
  \bibinfo{volume}{52} (\bibinfo{year}{2015}), \bibinfo{pages}{523--532}.
\newblock


\bibitem[\protect\citeauthoryear{Gonzalez-Franco and Lanier}{Gonzalez-Franco
  and Lanier}{2017}]%
        {gonzalez2017model}
\bibfield{author}{\bibinfo{person}{Mar Gonzalez-Franco} {and}
  \bibinfo{person}{Jaron Lanier}.} \bibinfo{year}{2017}\natexlab{}.
\newblock \showarticletitle{Model of Illusions and Virtual Reality}.
\newblock \bibinfo{journal}{\emph{Frontiers in psychology}}
  \bibinfo{volume}{8} (\bibinfo{year}{2017}).
\newblock


\bibitem[\protect\citeauthoryear{Gordon, Bigham, and Lasecki}{Gordon
  et~al\mbox{.}}{2015}]%
        {gordon2015legiontools}
\bibfield{author}{\bibinfo{person}{Mitchell Gordon},
  \bibinfo{person}{Jeffrey~P. Bigham}, {and} \bibinfo{person}{Walter~S.
  Lasecki}.} \bibinfo{year}{2015}\natexlab{}.
\newblock \showarticletitle{LegionTools: A Toolkit + UI for Recruiting and
  Routing Crowds to Synchronous Real-Time Tasks}. In
  \bibinfo{booktitle}{\emph{Adjunct Proceedings of the 28th Annual ACM
  Symposium on User Interface Software \& Technology}}
  \emph{(\bibinfo{series}{UIST '15 Adjunct})}. \bibinfo{publisher}{ACM},
  \bibinfo{address}{New York, NY, USA}, \bibinfo{pages}{81--82}.
\newblock
\showISBNx{978-1-4503-3780-9}


\bibitem[\protect\citeauthoryear{Hartig, Mang, and Evans}{Hartig
  et~al\mbox{.}}{1991}]%
        {hartig1991restorative}
\bibfield{author}{\bibinfo{person}{Terry Hartig}, \bibinfo{person}{Marlis
  Mang}, {and} \bibinfo{person}{Gary~W Evans}.}
  \bibinfo{year}{1991}\natexlab{}.
\newblock \showarticletitle{Restorative effects of natural environment
  experiences}.
\newblock \bibinfo{journal}{\emph{Environment and behavior}}
  \bibinfo{volume}{23}, \bibinfo{number}{1} (\bibinfo{year}{1991}),
  \bibinfo{pages}{3--26}.
\newblock


\bibitem[\protect\citeauthoryear{Hock, Benedikter, Gugenheimer, and
  Rukzio}{Hock et~al\mbox{.}}{2017}]%
        {hock2017carvr}
\bibfield{author}{\bibinfo{person}{Philipp Hock}, \bibinfo{person}{Sebastian
  Benedikter}, \bibinfo{person}{Jan Gugenheimer}, {and} \bibinfo{person}{Enrico
  Rukzio}.} \bibinfo{year}{2017}\natexlab{}.
\newblock \showarticletitle{CarVR: Enabling In-Car Virtual Reality
  Entertainment}. In \bibinfo{booktitle}{\emph{Proceedings of the 2017 CHI
  Conference on Human Factors in Computing Systems}}. ACM,
  \bibinfo{pages}{4034--4044}.
\newblock


\bibitem[\protect\citeauthoryear{Horton, Rand, and Zeckhauser}{Horton
  et~al\mbox{.}}{2011}]%
        {horton2011online}
\bibfield{author}{\bibinfo{person}{John~J Horton}, \bibinfo{person}{David~G
  Rand}, {and} \bibinfo{person}{Richard~J Zeckhauser}.}
  \bibinfo{year}{2011}\natexlab{}.
\newblock \showarticletitle{The online laboratory: Conducting experiments in a
  real labor market}.
\newblock \bibinfo{journal}{\emph{Experimental Economics}}
  \bibinfo{volume}{14}, \bibinfo{number}{3} (\bibinfo{year}{2011}),
  \bibinfo{pages}{399--425}.
\newblock


\bibitem[\protect\citeauthoryear{Huff and Tingley}{Huff and Tingley}{2015}]%
        {huff2015these}
\bibfield{author}{\bibinfo{person}{Connor Huff} {and} \bibinfo{person}{Dustin
  Tingley}.} \bibinfo{year}{2015}\natexlab{}.
\newblock \showarticletitle{``Who are these people?'' Evaluating the
  demographic characteristics and political preferences of MTurk survey
  respondents}.
\newblock \bibinfo{journal}{\emph{Research \& Politics}} \bibinfo{volume}{2},
  \bibinfo{number}{3} (\bibinfo{year}{2015}).
\newblock


\bibitem[\protect\citeauthoryear{Ipeirotis}{Ipeirotis}{2010}]%
        {ipeirotis2010demographics}
\bibfield{author}{\bibinfo{person}{Panagiotis~G Ipeirotis}.}
  \bibinfo{year}{2010}\natexlab{}.
\newblock \showarticletitle{Demographics of mechanical turk}.
\newblock \bibinfo{journal}{\emph{NYU Working Paper No. CEDER-10-01}}
  (\bibinfo{year}{2010}).
\newblock


\bibitem[\protect\citeauthoryear{Kamen}{Kamen}{2017}]%
        {kamen2017samsung}
\bibfield{author}{\bibinfo{person}{Matt Kamen}.}
  \bibinfo{year}{2017}\natexlab{}.
\newblock \bibinfo{title}{Samsung Gear VR shipped more devices than Oculus, HTC
  Vive, and PSVR combined}.
\newblock   (\bibinfo{date}{Feb} \bibinfo{year}{2017}).
\newblock
\urldef\tempurl%
\url{http://www.wired.co.uk/article/samsung-vr-outships-psvr-htc-vive-and-oculus}
\showURL{%
\tempurl}
\newblock
\shownote{Accessed on Oct 2017.}


\bibitem[\protect\citeauthoryear{Karmanov and Hamel}{Karmanov and
  Hamel}{2008}]%
        {karmanov2008assessing}
\bibfield{author}{\bibinfo{person}{Dmitri Karmanov} {and}
  \bibinfo{person}{Ronald Hamel}.} \bibinfo{year}{2008}\natexlab{}.
\newblock \showarticletitle{Assessing the restorative potential of contemporary
  urban environment (s): Beyond the nature versus urban dichotomy}.
\newblock \bibinfo{journal}{\emph{Landscape and Urban Planning}}
  \bibinfo{volume}{86}, \bibinfo{number}{2} (\bibinfo{year}{2008}),
  \bibinfo{pages}{115--125}.
\newblock


\bibitem[\protect\citeauthoryear{Kennedy, Lane, Berbaum, and
  Lilienthal}{Kennedy et~al\mbox{.}}{1993}]%
        {kennedy1993simulator}
\bibfield{author}{\bibinfo{person}{Robert~S Kennedy}, \bibinfo{person}{Norman~E
  Lane}, \bibinfo{person}{Kevin~S Berbaum}, {and} \bibinfo{person}{Michael~G
  Lilienthal}.} \bibinfo{year}{1993}\natexlab{}.
\newblock \showarticletitle{Simulator sickness questionnaire: An enhanced
  method for quantifying simulator sickness}.
\newblock \bibinfo{journal}{\emph{The international journal of aviation
  psychology}} \bibinfo{volume}{3}, \bibinfo{number}{3} (\bibinfo{year}{1993}),
  \bibinfo{pages}{203--220}.
\newblock


\bibitem[\protect\citeauthoryear{Kim, Sterman, Cohen, and Bernstein}{Kim
  et~al\mbox{.}}{2017}]%
        {kim2017mechanical}
\bibfield{author}{\bibinfo{person}{Joy Kim}, \bibinfo{person}{Sarah Sterman},
  \bibinfo{person}{Allegra Argent~Beal Cohen}, {and}
  \bibinfo{person}{Michael~S. Bernstein}.} \bibinfo{year}{2017}\natexlab{}.
\newblock \showarticletitle{Mechanical Novel: Crowdsourcing Complex Work
  Through Reflection and Revision}. In \bibinfo{booktitle}{\emph{Proceedings of
  the 2017 ACM Conference on Computer Supported Cooperative Work and Social
  Computing}} \emph{(\bibinfo{series}{CSCW '17})}. \bibinfo{publisher}{ACM},
  \bibinfo{address}{New York, NY, USA}, \bibinfo{pages}{233--245}.
\newblock
\showISBNx{978-1-4503-4335-0}
\urldef\tempurl%
\url{https://doi.org/10.1145/2998181.2998196}
\showDOI{\tempurl}


\bibitem[\protect\citeauthoryear{Kraut, Olson, Banaji, Bruckman, Cohen, and
  Couper}{Kraut et~al\mbox{.}}{2004}]%
        {kraut2004psychological}
\bibfield{author}{\bibinfo{person}{Robert Kraut}, \bibinfo{person}{Judith
  Olson}, \bibinfo{person}{Mahzarin Banaji}, \bibinfo{person}{Amy Bruckman},
  \bibinfo{person}{Jeffrey Cohen}, {and} \bibinfo{person}{Mick Couper}.}
  \bibinfo{year}{2004}\natexlab{}.
\newblock \showarticletitle{Psychological research online: report of Board of
  Scientific Affairs' Advisory Group on the Conduct of Research on the
  Internet.}
\newblock \bibinfo{journal}{\emph{American psychologist}} \bibinfo{volume}{59},
  \bibinfo{number}{2} (\bibinfo{year}{2004}), \bibinfo{pages}{105}.
\newblock


\bibitem[\protect\citeauthoryear{Lasecki, Murray, White, Miller, and
  Bigham}{Lasecki et~al\mbox{.}}{2011}]%
        {lasecki2011real}
\bibfield{author}{\bibinfo{person}{Walter~S. Lasecki}, \bibinfo{person}{Kyle~I.
  Murray}, \bibinfo{person}{Samuel White}, \bibinfo{person}{Robert~C. Miller},
  {and} \bibinfo{person}{Jeffrey~P. Bigham}.} \bibinfo{year}{2011}\natexlab{}.
\newblock \showarticletitle{Real-time Crowd Control of Existing Interfaces}. In
  \bibinfo{booktitle}{\emph{Proceedings of the 24th Annual ACM Symposium on
  User Interface Software and Technology}} \emph{(\bibinfo{series}{UIST '11})}.
  \bibinfo{publisher}{ACM}, \bibinfo{address}{New York, NY, USA},
  \bibinfo{pages}{23--32}.
\newblock
\showISBNx{978-1-4503-0716-1}


\bibitem[\protect\citeauthoryear{Lee, Kim, Daher, Raij, Schubert, Bailenson,
  and Welch}{Lee et~al\mbox{.}}{2016}]%
        {lee2016wobbly}
\bibfield{author}{\bibinfo{person}{Myungho Lee}, \bibinfo{person}{Kangsoo Kim},
  \bibinfo{person}{Salam Daher}, \bibinfo{person}{Andrew Raij},
  \bibinfo{person}{Ryan Schubert}, \bibinfo{person}{Jeremy Bailenson}, {and}
  \bibinfo{person}{Greg Welch}.} \bibinfo{year}{2016}\natexlab{}.
\newblock \showarticletitle{The wobbly table: Increased social presence via
  subtle incidental movement of a real-virtual table}. In
  \bibinfo{booktitle}{\emph{Virtual Reality (VR), 2016 IEEE}}. IEEE,
  \bibinfo{pages}{11--17}.
\newblock


\bibitem[\protect\citeauthoryear{Little, Chilton, Goldman, and Miller}{Little
  et~al\mbox{.}}{2010}]%
        {little2010turkit}
\bibfield{author}{\bibinfo{person}{Greg Little}, \bibinfo{person}{Lydia~B
  Chilton}, \bibinfo{person}{Max Goldman}, {and} \bibinfo{person}{Robert~C
  Miller}.} \bibinfo{year}{2010}\natexlab{}.
\newblock \showarticletitle{Turkit: human computation algorithms on mechanical
  turk}. In \bibinfo{booktitle}{\emph{Proceedings of the 23nd annual ACM
  symposium on User interface software and technology}}. ACM,
  \bibinfo{pages}{57--66}.
\newblock


\bibitem[\protect\citeauthoryear{Maani, Hoffman, Morrow, Maiers, Gaylord,
  McGhee, and DeSocio}{Maani et~al\mbox{.}}{2011}]%
        {maani2011virtual}
\bibfield{author}{\bibinfo{person}{Christopher~V Maani},
  \bibinfo{person}{Hunter~G Hoffman}, \bibinfo{person}{Michelle Morrow},
  \bibinfo{person}{Alan Maiers}, \bibinfo{person}{Kathryn Gaylord},
  \bibinfo{person}{Laura~L McGhee}, {and} \bibinfo{person}{Peter~A DeSocio}.}
  \bibinfo{year}{2011}\natexlab{}.
\newblock \showarticletitle{Virtual reality pain control during burn wound
  debridement of combat-related burn injuries using robot-like arm mounted VR
  goggles}.
\newblock \bibinfo{journal}{\emph{The Journal of trauma}} \bibinfo{volume}{71},
  \bibinfo{number}{1 0} (\bibinfo{year}{2011}), \bibinfo{pages}{S125}.
\newblock


\bibitem[\protect\citeauthoryear{Mao, Chen, Gajos, Parkes, Procaccia, and
  Zhang}{Mao et~al\mbox{.}}{2012}]%
        {mao2012turkserver}
\bibfield{author}{\bibinfo{person}{Andrew Mao}, \bibinfo{person}{Yiling Chen},
  \bibinfo{person}{Krzysztof~Z Gajos}, \bibinfo{person}{David Parkes},
  \bibinfo{person}{Ariel~D Procaccia}, {and} \bibinfo{person}{Haoqi Zhang}.}
  \bibinfo{year}{2012}\natexlab{}.
\newblock \showarticletitle{Turkserver: Enabling synchronous and longitudinal
  online experiments}.
\newblock \bibinfo{journal}{\emph{Proceedings of HCOMP}}  \bibinfo{volume}{12}
  (\bibinfo{year}{2012}).
\newblock


\bibitem[\protect\citeauthoryear{Mao, Dworkin, Suri, and Watts}{Mao
  et~al\mbox{.}}{2017}]%
        {mao2017resilient}
\bibfield{author}{\bibinfo{person}{Andrew Mao}, \bibinfo{person}{Lili Dworkin},
  \bibinfo{person}{Siddharth Suri}, {and} \bibinfo{person}{Duncan~J Watts}.}
  \bibinfo{year}{2017}\natexlab{}.
\newblock \showarticletitle{Resilient cooperators stabilize long-run
  cooperation in the finitely repeated Prisoner’s Dilemma}.
\newblock \bibinfo{journal}{\emph{Nature communications}}  \bibinfo{volume}{8}
  (\bibinfo{year}{2017}).
\newblock


\bibitem[\protect\citeauthoryear{Mao, Mason, Suri, and Watts}{Mao
  et~al\mbox{.}}{2016}]%
        {mao2016experimental}
\bibfield{author}{\bibinfo{person}{Andrew Mao}, \bibinfo{person}{Winter Mason},
  \bibinfo{person}{Siddharth Suri}, {and} \bibinfo{person}{Duncan~J Watts}.}
  \bibinfo{year}{2016}\natexlab{}.
\newblock \showarticletitle{An experimental study of team size and performance
  on a complex task}.
\newblock \bibinfo{journal}{\emph{PloS one}} \bibinfo{volume}{11},
  \bibinfo{number}{4} (\bibinfo{year}{2016}).
\newblock


\bibitem[\protect\citeauthoryear{Mason and Suri}{Mason and Suri}{2012}]%
        {mason2012conducting}
\bibfield{author}{\bibinfo{person}{Winter Mason} {and}
  \bibinfo{person}{Siddharth Suri}.} \bibinfo{year}{2012}\natexlab{}.
\newblock \showarticletitle{Conducting behavioral research on Amazon's
  Mechanical Turk}.
\newblock \bibinfo{journal}{\emph{Behavior research methods}}
  \bibinfo{volume}{44}, \bibinfo{number}{1} (\bibinfo{year}{2012}),
  \bibinfo{pages}{1--23}.
\newblock


\bibitem[\protect\citeauthoryear{Milgram, Bickman, and Berkowitz}{Milgram
  et~al\mbox{.}}{1969}]%
        {milgram1969note}
\bibfield{author}{\bibinfo{person}{Stanley Milgram}, \bibinfo{person}{Leonard
  Bickman}, {and} \bibinfo{person}{Lawrence Berkowitz}.}
  \bibinfo{year}{1969}\natexlab{}.
\newblock \showarticletitle{Note on the drawing power of crowds of different
  size.}
\newblock \bibinfo{journal}{\emph{Journal of personality and social
  psychology}} \bibinfo{volume}{13}, \bibinfo{number}{2}
  (\bibinfo{year}{1969}), \bibinfo{pages}{79}.
\newblock


\bibitem[\protect\citeauthoryear{Nowak, Page, and Sigmund}{Nowak
  et~al\mbox{.}}{2000}]%
        {nowak2000fairness}
\bibfield{author}{\bibinfo{person}{Martin~A Nowak}, \bibinfo{person}{Karen~M
  Page}, {and} \bibinfo{person}{Karl Sigmund}.}
  \bibinfo{year}{2000}\natexlab{}.
\newblock \showarticletitle{Fairness versus reason in the ultimatum game}.
\newblock \bibinfo{journal}{\emph{Science}} \bibinfo{volume}{289},
  \bibinfo{number}{5485} (\bibinfo{year}{2000}), \bibinfo{pages}{1773--1775}.
\newblock


\bibitem[\protect\citeauthoryear{Oh, Bailenson, Kr{\"a}mer, and Li}{Oh
  et~al\mbox{.}}{2016a}]%
        {oh2016let}
\bibfield{author}{\bibinfo{person}{Soo~Youn Oh}, \bibinfo{person}{Jeremy
  Bailenson}, \bibinfo{person}{Nicole Kr{\"a}mer}, {and}
  \bibinfo{person}{Benjamin Li}.} \bibinfo{year}{2016}\natexlab{a}.
\newblock \showarticletitle{Let the Avatar Brighten Your Smile: Effects of
  Enhancing Facial Expressions in Virtual Environments}.
\newblock \bibinfo{journal}{\emph{PloS one}} \bibinfo{volume}{11},
  \bibinfo{number}{9} (\bibinfo{year}{2016}).
\newblock


\bibitem[\protect\citeauthoryear{Oh, Shriram, Laha, Baughman, Ogle, and
  Bailenson}{Oh et~al\mbox{.}}{2016b}]%
        {oh2016immersion}
\bibfield{author}{\bibinfo{person}{Soo~Youn Oh}, \bibinfo{person}{Ketaki
  Shriram}, \bibinfo{person}{Bireswar Laha}, \bibinfo{person}{Shawnee
  Baughman}, \bibinfo{person}{Elise Ogle}, {and} \bibinfo{person}{Jeremy
  Bailenson}.} \bibinfo{year}{2016}\natexlab{b}.
\newblock \showarticletitle{Immersion at scale: Researcher's guide to
  ecologically valid mobile experiments}. In \bibinfo{booktitle}{\emph{Virtual
  Reality (VR), 2016 IEEE}}. IEEE, \bibinfo{pages}{249--250}.
\newblock


\bibitem[\protect\citeauthoryear{Packer and Bond}{Packer and Bond}{2010}]%
        {packer2010museums}
\bibfield{author}{\bibinfo{person}{Jan Packer} {and} \bibinfo{person}{Nigel
  Bond}.} \bibinfo{year}{2010}\natexlab{}.
\newblock \showarticletitle{Museums as restorative environments}.
\newblock \bibinfo{journal}{\emph{Curator: The Museum Journal}}
  \bibinfo{volume}{53}, \bibinfo{number}{4} (\bibinfo{year}{2010}),
  \bibinfo{pages}{421--436}.
\newblock


\bibitem[\protect\citeauthoryear{Paolacci, Chandler, and Ipeirotis}{Paolacci
  et~al\mbox{.}}{2010}]%
        {paolacci2010running}
\bibfield{author}{\bibinfo{person}{Gabriele Paolacci}, \bibinfo{person}{Jesse
  Chandler}, {and} \bibinfo{person}{Panagiotis~G Ipeirotis}.}
  \bibinfo{year}{2010}\natexlab{}.
\newblock \showarticletitle{Running experiments on amazon mechanical turk (June
  24, 2010)}.
\newblock \bibinfo{journal}{\emph{Judgment and Decision Making, Vol. 5, No. 5,
  411-419.}} (\bibinfo{year}{2010}).
\newblock
\urldef\tempurl%
\url{https://ssrn.com/abstract=1626226}
\showURL{%
\tempurl}


\bibitem[\protect\citeauthoryear{Rawassizadeh, Khosravipour, and
  Tjoa}{Rawassizadeh et~al\mbox{.}}{2011}]%
        {rawassizadeh2011persuasive}
\bibfield{author}{\bibinfo{person}{Reza Rawassizadeh}, \bibinfo{person}{Soheil
  Khosravipour}, {and} \bibinfo{person}{A~Min Tjoa}.}
  \bibinfo{year}{2011}\natexlab{}.
\newblock \showarticletitle{A persuasive approach for indoor environment
  tidiness}.
\newblock \bibinfo{journal}{\emph{Sixth International Conference on Persuasive
  Technology}} (\bibinfo{year}{2011}).
\newblock


\bibitem[\protect\citeauthoryear{Reinecke and Gajos}{Reinecke and
  Gajos}{2015}]%
        {reinecke2015lab}
\bibfield{author}{\bibinfo{person}{Katharina Reinecke} {and}
  \bibinfo{person}{Krzysztof~Z. Gajos}.} \bibinfo{year}{2015}\natexlab{}.
\newblock \showarticletitle{LabintheWild: Conducting Large-Scale Online
  Experiments With Uncompensated Samples}. In
  \bibinfo{booktitle}{\emph{Proceedings of the 18th ACM Conference on Computer
  Supported Cooperative Work \& Social Computing}} \emph{(\bibinfo{series}{CSCW
  '15})}. \bibinfo{publisher}{ACM}, \bibinfo{address}{New York, NY, USA},
  \bibinfo{pages}{1364--1378}.
\newblock
\showISBNx{978-1-4503-2922-4}
\urldef\tempurl%
\url{https://doi.org/10.1145/2675133.2675246}
\showDOI{\tempurl}


\bibitem[\protect\citeauthoryear{Retelny, Robaszkiewicz, To, Lasecki, Patel,
  Rahmati, Doshi, Valentine, and Bernstein}{Retelny et~al\mbox{.}}{2014}]%
        {retelny2014expert}
\bibfield{author}{\bibinfo{person}{Daniela Retelny},
  \bibinfo{person}{S{\'e}bastien Robaszkiewicz}, \bibinfo{person}{Alexandra
  To}, \bibinfo{person}{Walter~S Lasecki}, \bibinfo{person}{Jay Patel},
  \bibinfo{person}{Negar Rahmati}, \bibinfo{person}{Tulsee Doshi},
  \bibinfo{person}{Melissa Valentine}, {and} \bibinfo{person}{Michael~S
  Bernstein}.} \bibinfo{year}{2014}\natexlab{}.
\newblock \showarticletitle{Expert crowdsourcing with flash teams}. In
  \bibinfo{booktitle}{\emph{Proceedings of the 27th annual ACM symposium on
  User interface software and technology}}. ACM, \bibinfo{pages}{75--85}.
\newblock


\bibitem[\protect\citeauthoryear{Robertson}{Robertson}{2017}]%
        {robertson2017google}
\bibfield{author}{\bibinfo{person}{Adi Robertson}.}
  \bibinfo{year}{2017}\natexlab{}.
\newblock \bibinfo{title}{Google has shipped over 10 million Cardboard VR
  headsets}.
\newblock   (\bibinfo{date}{Feb} \bibinfo{year}{2017}).
\newblock
\urldef\tempurl%
\url{https://www.theverge.com/2017/2/28/14767902/google-cardboard-10-million-shipped-vr-ar-apps}
\showURL{%
\tempurl}
\newblock
\shownote{Accessed on Oct 2017.}


\bibitem[\protect\citeauthoryear{Rosenberg, Baughman, and Bailenson}{Rosenberg
  et~al\mbox{.}}{2013}]%
        {rosenberg2013virtual}
\bibfield{author}{\bibinfo{person}{Robin~S. Rosenberg},
  \bibinfo{person}{Shawnee~L. Baughman}, {and} \bibinfo{person}{Jeremy~N.
  Bailenson}.} \bibinfo{year}{2013}\natexlab{}.
\newblock \showarticletitle{Virtual Superheroes: Using Superpowers in Virtual
  Reality to Encourage Prosocial Behavior}.
\newblock \bibinfo{journal}{\emph{PloS one}}  \bibinfo{volume}{8}
  (\bibinfo{date}{01} \bibinfo{year}{2013}).
\newblock


\bibitem[\protect\citeauthoryear{Salganik}{Salganik}{2017}]%
        {salganik2017bit}
\bibfield{author}{\bibinfo{person}{Matthew~J Salganik}.}
  \bibinfo{year}{2017}\natexlab{}.
\newblock \bibinfo{booktitle}{\emph{Bit by Bit: Social Research in the Digital
  Age}}.
\newblock \bibinfo{publisher}{Princeton University Press}.
\newblock


\bibitem[\protect\citeauthoryear{Schroeder}{Schroeder}{1996}]%
        {Schroeder1996}
\bibfield{author}{\bibinfo{person}{Ralph Schroeder}.}
  \bibinfo{year}{1996}\natexlab{}.
\newblock \bibinfo{booktitle}{\emph{Possible worlds : the social dynamic of
  virtual reality technology}}.
\newblock \bibinfo{publisher}{Westview Press}.
\newblock


\bibitem[\protect\citeauthoryear{Schwind, Knierim, Tasci, Franczak, Haas, and
  Henze}{Schwind et~al\mbox{.}}{2017}]%
        {schwind2017hands}
\bibfield{author}{\bibinfo{person}{Valentin Schwind}, \bibinfo{person}{Pascal
  Knierim}, \bibinfo{person}{Cagri Tasci}, \bibinfo{person}{Patrick Franczak},
  \bibinfo{person}{Nico Haas}, {and} \bibinfo{person}{Niels Henze}.}
  \bibinfo{year}{2017}\natexlab{}.
\newblock \showarticletitle{``These Are Not My Hands!'': Effect of Gender on
  the Perception of Avatar Hands in Virtual Reality}. In
  \bibinfo{booktitle}{\emph{Proceedings of the 2017 CHI Conference on Human
  Factors in Computing Systems}} \emph{(\bibinfo{series}{CHI '17})}.
  \bibinfo{publisher}{ACM}, \bibinfo{address}{New York, NY, USA},
  \bibinfo{pages}{1577--1582}.
\newblock
\showISBNx{978-1-4503-4655-9}
\urldef\tempurl%
\url{https://doi.org/10.1145/3025453.3025602}
\showDOI{\tempurl}


\bibitem[\protect\citeauthoryear{Sheng, Provost, and Ipeirotis}{Sheng
  et~al\mbox{.}}{2008}]%
        {sheng2008get}
\bibfield{author}{\bibinfo{person}{Victor~S. Sheng}, \bibinfo{person}{Foster
  Provost}, {and} \bibinfo{person}{Panagiotis~G. Ipeirotis}.}
  \bibinfo{year}{2008}\natexlab{}.
\newblock \showarticletitle{Get Another Label? Improving Data Quality and Data
  Mining Using Multiple, Noisy Labelers}. In
  \bibinfo{booktitle}{\emph{Proceedings of the 14th ACM SIGKDD International
  Conference on Knowledge Discovery and Data Mining}}
  \emph{(\bibinfo{series}{KDD '08})}. \bibinfo{publisher}{ACM},
  \bibinfo{address}{New York, NY, USA}, \bibinfo{pages}{614--622}.
\newblock
\showISBNx{978-1-60558-193-4}


\bibitem[\protect\citeauthoryear{Shriram and Schwartz}{Shriram and
  Schwartz}{2017}]%
        {shriram2017all}
\bibfield{author}{\bibinfo{person}{Ketaki Shriram} {and} \bibinfo{person}{Raz
  Schwartz}.} \bibinfo{year}{2017}\natexlab{}.
\newblock \showarticletitle{All are welcome: Using VR ethnography to explore
  harassment behavior in immersive social virtual reality}. In
  \bibinfo{booktitle}{\emph{Virtual Reality (VR), 2017 IEEE}}. IEEE,
  \bibinfo{pages}{225--226}.
\newblock


\bibitem[\protect\citeauthoryear{Sitton}{Sitton}{1984}]%
        {sitton1984messy}
\bibfield{author}{\bibinfo{person}{Sarah Sitton}.}
  \bibinfo{year}{1984}\natexlab{}.
\newblock \showarticletitle{The messy desk effect: How tidiness affects the
  perception of others}.
\newblock \bibinfo{journal}{\emph{The Journal of psychology}}
  \bibinfo{volume}{117}, \bibinfo{number}{2} (\bibinfo{year}{1984}),
  \bibinfo{pages}{263--267}.
\newblock


\bibitem[\protect\citeauthoryear{Slater}{Slater}{2009}]%
        {slater2009place}
\bibfield{author}{\bibinfo{person}{Mel Slater}.}
  \bibinfo{year}{2009}\natexlab{}.
\newblock \showarticletitle{Place illusion and plausibility can lead to
  realistic behaviour in immersive virtual environments}.
\newblock \bibinfo{journal}{\emph{Philosophical Transactions of the Royal
  Society of London B: Biological Sciences}} \bibinfo{volume}{364},
  \bibinfo{number}{1535} (\bibinfo{year}{2009}), \bibinfo{pages}{3549--3557}.
\newblock


\bibitem[\protect\citeauthoryear{Suri and Watts}{Suri and Watts}{2011}]%
        {suri2011cooperation}
\bibfield{author}{\bibinfo{person}{Siddharth Suri} {and}
  \bibinfo{person}{Duncan~J Watts}.} \bibinfo{year}{2011}\natexlab{}.
\newblock \showarticletitle{Cooperation and contagion in web-based, networked
  public goods experiments}.
\newblock \bibinfo{journal}{\emph{PloS one}} \bibinfo{volume}{6},
  \bibinfo{number}{3} (\bibinfo{year}{2011}), \bibinfo{pages}{e16836}.
\newblock


\bibitem[\protect\citeauthoryear{Ulrich}{Ulrich}{1981}]%
        {ulrich1981natural}
\bibfield{author}{\bibinfo{person}{Roger~S Ulrich}.}
  \bibinfo{year}{1981}\natexlab{}.
\newblock \showarticletitle{Natural versus urban scenes: Some
  psychophysiological effects}.
\newblock \bibinfo{journal}{\emph{Environment and behavior}}
  \bibinfo{volume}{13}, \bibinfo{number}{5} (\bibinfo{year}{1981}),
  \bibinfo{pages}{523--556}.
\newblock


\bibitem[\protect\citeauthoryear{Ulrich, Simons, Losito, Fiorito, Miles, and
  Zelson}{Ulrich et~al\mbox{.}}{1991}]%
        {ulrich1991stress}
\bibfield{author}{\bibinfo{person}{Roger~S Ulrich}, \bibinfo{person}{Robert~F
  Simons}, \bibinfo{person}{Barbara~D Losito}, \bibinfo{person}{Evelyn
  Fiorito}, \bibinfo{person}{Mark~A Miles}, {and} \bibinfo{person}{Michael
  Zelson}.} \bibinfo{year}{1991}\natexlab{}.
\newblock \showarticletitle{Stress recovery during exposure to natural and
  urban environments}.
\newblock \bibinfo{journal}{\emph{Journal of environmental psychology}}
  \bibinfo{volume}{11}, \bibinfo{number}{3} (\bibinfo{year}{1991}),
  \bibinfo{pages}{201--230}.
\newblock


\bibitem[\protect\citeauthoryear{Valtchanov, Barton, and Ellard}{Valtchanov
  et~al\mbox{.}}{2010}]%
        {valtchanov2010restorative}
\bibfield{author}{\bibinfo{person}{Deltcho Valtchanov},
  \bibinfo{person}{Kevin~R Barton}, {and} \bibinfo{person}{Colin Ellard}.}
  \bibinfo{year}{2010}\natexlab{}.
\newblock \showarticletitle{Restorative effects of virtual nature settings}.
\newblock \bibinfo{journal}{\emph{Cyberpsychology, Behavior, and Social
  Networking}} \bibinfo{volume}{13}, \bibinfo{number}{5}
  (\bibinfo{year}{2010}), \bibinfo{pages}{503--512}.
\newblock


\bibitem[\protect\citeauthoryear{Van Der~Heide, Schumaker, Peterson, and
  Jones}{Van Der~Heide et~al\mbox{.}}{2013}]%
        {van2013proteus}
\bibfield{author}{\bibinfo{person}{Brandon Van Der~Heide},
  \bibinfo{person}{Erin~M Schumaker}, \bibinfo{person}{Ashley~M Peterson},
  {and} \bibinfo{person}{Elizabeth~B Jones}.} \bibinfo{year}{2013}\natexlab{}.
\newblock \showarticletitle{The Proteus effect in dyadic communication:
  Examining the effect of avatar appearance in computer-mediated dyadic
  interaction}.
\newblock \bibinfo{journal}{\emph{Communication Research}}
  \bibinfo{volume}{40}, \bibinfo{number}{6} (\bibinfo{year}{2013}),
  \bibinfo{pages}{838--860}.
\newblock


\bibitem[\protect\citeauthoryear{Vohs, Redden, and Rahinel}{Vohs
  et~al\mbox{.}}{2013}]%
        {vohs2013physical}
\bibfield{author}{\bibinfo{person}{Kathleen~D Vohs}, \bibinfo{person}{Joseph~P
  Redden}, {and} \bibinfo{person}{Ryan Rahinel}.}
  \bibinfo{year}{2013}\natexlab{}.
\newblock \showarticletitle{Physical order produces healthy choices,
  generosity, and conventionality, whereas disorder produces creativity}.
\newblock \bibinfo{journal}{\emph{Psychological Science}} \bibinfo{volume}{24},
  \bibinfo{number}{9} (\bibinfo{year}{2013}), \bibinfo{pages}{1860--1867}.
\newblock


\bibitem[\protect\citeauthoryear{Witmer and Singer}{Witmer and Singer}{1998}]%
        {witmer1998measuring}
\bibfield{author}{\bibinfo{person}{Bob~G Witmer} {and}
  \bibinfo{person}{Michael~J Singer}.} \bibinfo{year}{1998}\natexlab{}.
\newblock \showarticletitle{Measuring presence in virtual environments: A
  presence questionnaire}.
\newblock \bibinfo{journal}{\emph{Presence: Teleoperators and virtual
  environments}} \bibinfo{volume}{7}, \bibinfo{number}{3}
  (\bibinfo{year}{1998}), \bibinfo{pages}{225--240}.
\newblock


\bibitem[\protect\citeauthoryear{Won, Bailenson, Lee, and Lanier}{Won
  et~al\mbox{.}}{2015}]%
        {won2015homuncular}
\bibfield{author}{\bibinfo{person}{Andrea~Stevenson Won},
  \bibinfo{person}{Jeremy Bailenson}, \bibinfo{person}{Jimmy Lee}, {and}
  \bibinfo{person}{Jaron Lanier}.} \bibinfo{year}{2015}\natexlab{}.
\newblock \showarticletitle{Homuncular flexibility in virtual reality}.
\newblock \bibinfo{journal}{\emph{Journal of Computer-Mediated Communication}}
  \bibinfo{volume}{20}, \bibinfo{number}{3} (\bibinfo{year}{2015}),
  \bibinfo{pages}{241--259}.
\newblock


\bibitem[\protect\citeauthoryear{Yee and Bailenson}{Yee and Bailenson}{2006}]%
        {yee2006walk}
\bibfield{author}{\bibinfo{person}{Nick Yee} {and} \bibinfo{person}{Jeremy
  Bailenson}.} \bibinfo{year}{2006}\natexlab{}.
\newblock \showarticletitle{Walk a mile in digital shoes: The impact of
  embodied perspective-taking on the reduction of negative stereotyping in
  immersive virtual environments}.
\newblock \bibinfo{journal}{\emph{Presence Teleoperators \& Virtual
  Environments}} (\bibinfo{date}{01} \bibinfo{year}{2006}).
\newblock


\bibitem[\protect\citeauthoryear{Yee and Bailenson}{Yee and Bailenson}{2007}]%
        {yee2007proteus}
\bibfield{author}{\bibinfo{person}{Nick Yee} {and} \bibinfo{person}{Jeremy
  Bailenson}.} \bibinfo{year}{2007}\natexlab{}.
\newblock \showarticletitle{The Proteus effect: The effect of transformed
  self-representation on behavior}.
\newblock \bibinfo{journal}{\emph{Human communication research}}
  \bibinfo{volume}{33}, \bibinfo{number}{3} (\bibinfo{year}{2007}),
  \bibinfo{pages}{271--290}.
\newblock


\bibitem[\protect\citeauthoryear{Yee, Bailenson, and Ducheneaut}{Yee
  et~al\mbox{.}}{2009}]%
        {yee2009proteus}
\bibfield{author}{\bibinfo{person}{Nick Yee}, \bibinfo{person}{Jeremy~N
  Bailenson}, {and} \bibinfo{person}{Nicolas Ducheneaut}.}
  \bibinfo{year}{2009}\natexlab{}.
\newblock \showarticletitle{The Proteus effect: Implications of transformed
  digital self-representation on online and offline behavior}.
\newblock \bibinfo{journal}{\emph{Communication Research}}
  \bibinfo{volume}{36}, \bibinfo{number}{2} (\bibinfo{year}{2009}),
  \bibinfo{pages}{285--312}.
\newblock


\bibitem[\protect\citeauthoryear{Zuckerman}{Zuckerman}{1977}]%
        {zuckerman1977development}
\bibfield{author}{\bibinfo{person}{Marvin Zuckerman}.}
  \bibinfo{year}{1977}\natexlab{}.
\newblock \showarticletitle{Development of a situation-specific trait-state
  test for the prediction and measurement of affective responses.}
\newblock \bibinfo{journal}{\emph{Journal of Consulting and clinical
  psychology}} \bibinfo{volume}{45}, \bibinfo{number}{4}
  (\bibinfo{year}{1977}), \bibinfo{pages}{513}.
\newblock


\end{thebibliography}

\end{document}